\documentclass[12pt]{article}
\usepackage{graphicx} % Required for inserting images
\usepackage[utf8]{inputenc} % For UTF-8 encoding
\usepackage{amsmath, amssymb} % Math symbols
\usepackage[margin=1in]{geometry} % Adjust page margins
\usepackage{hyperref}

\usepackage{authblk}
\usepackage{subcaption}
\usepackage{array}
\usepackage{placeins}
\usepackage{multirow}
\usepackage{multicol}
\usepackage{cite}

\title{Demonstration of Quantum-Secure Communications in a Nuclear Reactor}

\author[1]{Konstantinos Gkouliaras \footnote{kgkoulia@purdue.edu}}
\author[1]{Vasileios Theos}
\author[1]{True Miller}
\author[1]{Brian Jowers}
\author[2]{George Kennedy}
\author[2]{Andy Grant}
\author[3]{Terry Cronin}
\author[4]{Philip G. Evans}
\author[1]{Stylianos Chatzidakis}
\affil[1]{School of Nuclear Engineering, Purdue University, USA}
\affil[2]{Toshiba Europe Ltd, UK}
\affil[3]{Toshiba International Corporation, USA}
\affil[4]{Computational Sciences and Engineering Division, Oak Ridge National Laboratory, USA}

\date{May 2025}
\begin{document}

\maketitle

\begin{abstract}

Quantum key distribution (QKD), one of the latest cryptographic techniques, founded on the laws of quantum mechanics rather than mathematical complexity, promises for the first time unconditional secure remote communications. Integrating this technology into the next generation nuclear systems - designed for universal data collection and real-time sharing as well as cutting-edge instrumentation and increased dependency on digital technologies - could provide significant benefits enabling secure, unattended, and autonomous operation in remote areas, e.g., microreactors and fission batteries. However, any practical implementation on a critical reactor system must meet strict requirements on latency, control system compatibility, stability, and performance under operational transients. Here, we report the  complete end-to-end demonstration of a phase-encoding decoy-state BB84 protocol QKD system under prototypic conditions on Purdue’s fully digital nuclear reactor, PUR-1. The system was installed in PUR-1 successfully executing real-time encryption and decryption of 2,000 signals over optic fiber distances up to 82 km using OTP-based encryption and up to 140 km with AES-based encryption. For a core of 68 signals, OTP-secure communication was achieved for up to 135 km. The QKD system maintained a stable secret key rate of 320 kbps and a quantum bit error of 3.8\% at 54 km. Our results demonstrate that OTP-based encryption introduces minimal latency while the more key-efficient AES and ASCON encryption schemes can significantly increase the number of signals encrypted without latency penalties. Additionally, implementation of a dynamic key pool ensures several hours of secure key availability during potential system downtimes. This work shows the potential of quantum-based secure remote communications for future digitally driven nuclear reactor technologies.

\end{abstract}

\section{Introduction}

Advanced reactor designs (e.g., Microreactors, fission batteries) are proposed with unique new capabilities, such as remote monitoring and semi-autonomous operation. Their development is characterized by the goals of minimizing economic cost, leveraging passive safety systems, and enabling support of different energy outlet types \cite{fasano_advance_2021}. Notably, advanced reactor designs would allow for a wide range of applications complementary to electricity generation, such as district heating and water desalination, while being compatible with grid-connected and off-grid applications. As of today, seventy Small Modular Reactor (SMR) designs have been reported, while three of them are currently operational. Of those, forty designs are advanced Gen-IV reactors \cite{noland_overview_2025}.

While this architecture offers numerous advantages, it is nevertheless vulnerable to cyber attacks. The “cyber-attack surface” (i.e., the potentially vulnerable systems and components) is greater compared to conventional reactor designs, as a result of increased automation, I\&C complexity, remote operation and operational personnel reduction \cite{fasano_advance_2021}. 
The nuclear industry has been a target of such cyber attacks for the past 30 years, with many attacks aimed at gaining intelligence on Supervisory Control And Data Acquisition (SCADA) networks. This was ably demonstrated in attacks launched in 2014 against the Korea Hydro and Nuclear Power Co. that specifically targeted the blueprints and electrical flow charts of nuclear reactors \cite{baylon_cyber_2015}. Independent of the objectives or aims of an attack, attackers almost always first gather information about the target system to identify network topology, software versions, authorization or authentication mechanisms, and critical targets. This highlights that the first critical layer of defense against attacks would be to guarantee the confidentiality and authentication of any communication.

While cryptographic implementation was first proposed by the US NRC in 2010 \cite{nrc_regulatory_2010}, practical implementation at an actual power plant has not yet taken place \cite{son_modeling_2023}.
Information-theoretical security (i.e., unconditional) can be achieved when the well-known symmetric encryption One-Time Pad (OTP) algorithm is implemented. To do so, OTP requires that communication parties have access to continuously refreshed and truly random keys, equal in size to the encrypted data \cite{katz_introduction_2021}. These requirements are not easily implemented, as never-used keys need to be distributed between two parties in real time. Instead, current cryptographic schemes are based on the computational complexity of public key cryptography, i.e., on the difficulty of reversing certain mathematical functions. Unfortunately, public key cryptography has been shown to be vulnerable to the advent of quantum computers and Shor’s algorithm \cite{mavroeidis_impact_2018, gidney_how_2021, szikora_end_2022}. For instance, predictions on the evolution and scalability of quantum computers reveal that they could compromise public key encryption (RSA) within hours \cite{azhari_analyzing_2024}. Large scale quantum computers in the next decade is a realistic expectation with several initiatives launched, including the National Quantum Initiative Act \cite{noauthor_national_nodate}, Google’s AI Quantum Laboratory with plans to commercialize quantum computers \cite{neven_meet_2024,porter_google_2021, roth_google_2024}, IBM Q \cite{murphy_ai_2025, abughanem_ibm_2025}, etc.

Two main approaches are currently being explored to address this challenge: Post-Quantum Cryptography (PQC) and Quantum Cryptography (QC). PQC, an active area of research led by NIST focusing on quantum-safe algorithms \cite{joseph_transitioning_2022}, aims to develop classical cryptographic schemes which do not offer quantum computers an advantage over classical ones. QKD relies on the inherent properties of quantum mechanics to deliver unconditional security, without making assumptions on adversarial resources or strategy.

QKD is the most mature quantum cryptography application, enabling secure generation and distribution of a truly random key at two distant locations \cite{brazaola-vicario_quantum_2024}. Leveraging the laws of quantum physics, specifically the uncertainty principle and the no-cloning theorem, QKD has been extensively studied. Since its first introduction in 1984 with the BB84 protocol \cite{bennett_quantum_2014}, there have been multiple protocols \cite{nurhadi_quantum_2018,wei_decoy-state_2013, lucamarini_overcoming_2018}, security proofs \cite{pirandola_advances_2020, portmann_security_2022, tomamichel_tight_2012, yin_security_2016, pirandola_fundamental_2017, takeoka_fundamental_2014, li_improving_2022, chen_continuous-mode_2023}, hardware advancements \cite{zhang_advances_2015, grunenfelder_fast_2023}, experimental network demonstrations \cite{cao_evolution_2022, kurtsiefer_step_2002, ursin_entanglement-based_2007, gerhardt_full-field_2011, chen_twin-field_2021, chen_integrated_2021, pittaluga_600-km_2021}, and commercial realizations \cite{balci_strategic_2020}.

This paper presents an experimental demonstration of quantum-secure remote monitoring of a real-world nuclear reactor. Coupling the TOSHIBA Long-Distance QKD system (QKD-LD) with Purdue University's nuclear Reactor number One (PUR-1), we assemble a testbed for real-time exchange of encrypted reactor data. A series of measurements are conducted to evaluate key generation rates and system latency under different transmission distances, operational use cases, and encryption algorithms. Domain knowledge and experimental measurements are used to investigate the configuration-specific conditions for maintaining secure communication, even in the scenario of a QKD failure.

The significance of our implementation stems from two main reasons: firstly, it narrows the gap between theoretical/numerical estimations and real-world applications. The presented setup generates hands-on performance metrics and replicates the full Nuclear Power Plant (NPP) communication cycle (data generation, encryption, transmission, decryption, and storage). Secondly, it permits one to study in detail the nature and requirements of nuclear reactor data generated from an actual power plant. The study does not evaluate system performance generically from a secure network perspective, but incorporates domain knowledge originating from physics constraints, operational constraints, official regulation, and more.

The structure of this paper is as follows. \autoref{sec:qkd} features a brief overview of QKD fundamentals, and \autoref{sec:related_work} provides a background of nuclear cybersecurity and QKD applications for critical infrastructure. \autoref{sec:communication_model} introduces a description and mathematical formulation of the secure communication model. In \autoref{sec:pur1}, the PUR-1 nuclear reactor facility is presented and high-level operational use cases are defined. \autoref{sec:experimental} analyzes the experimental setup installed at PUR-1, while  \autoref{sec:secret_key_generation} and \autoref{sec:latency} demonstrate the evaluated use-case configurations and discuss the corresponding results, with respect to key availability and latency, respectively. Finally, \autoref{sec:conclusion} presents concluding remarks and directions regarding the expansion of the present work.

\section{Quantum Key Distribution}
\label{sec:qkd}

QKD was first introduced in 1984 by Bennett and Brassard \cite{bennett_quantum_2014}. A physical-layer security scheme, it leverages the laws of quantum mechanics to deliver information-theoretic security. QKD eventually generates truly random keys and delivers them in real time to the communication parties. A generic QKD system consists of two channels: a quantum channel (optical fiber or free space) and a classical channel (data link). The quantum channel is used to exchange single photons which carry information encoded in one of their degrees of freedom, such as polarization or phase.  Once the photon pulses are detected, a series of communication rounds occur via the classical channel to correct errors and remove any knowledge potentially leaked to an adversary. The two parties gain access to quantum-random bit strings, which can be used with symmetric encryption to provide unconditional security \cite{djordjevic_physical-layer_2019, wolf_quantum_2021}.

The security guarantee of QKD lies in the principles of quantum physics. Due to the no-cloning theorem and Heisenberg's uncertainty principle, a potential adversary attempting to intercept the photons would cause the collapse of the quantum state. An adversarial attempt to recreate it would inevitably lead to the introduction of errors, which would be detectable by the legitimate communication parties. As a result, QKD-enhanced encryption does not rely on adversarial computational power or strategy, providing future-proof security even against a quantum-computer-initiated attack. In-depth discussions of QKD security for practical settings can be found in \cite{scarani_security_2009} and \cite{portmann_security_2022}.

The main metric to evaluate QKD performance is the Secret Key Rate (SKR), describing the frequency at which secure bit streams are generated and become available to the parties. Abstractly speaking, SKR can be traced back to the raw key rate $R_{\text{raw}}$ and the Quantum Bit Error Rate (QBER). The raw key is equal in size to the number of pulses detected at the receiver, representing the bit streams before any classical processing has taken place (sifting, error correction, privacy amplification). It is primarily affected by channel attenuation, a quantity proportional to the transmission distance, as well as by hardware imperfections (detector efficiency, source repetition rate, etc.). The sifted key is formed by discarding measurements conducted using a different basis than the one used for encoding. The key reduction in this stage is based on the selected protocol and is modeled with $\eta_{\text{sift}}$. 

\begin{equation}
    \eta_{\text{sift}}\triangleq\frac{\text{size of sifted key}}{\text{size of raw key}}
    \label{eq:qber_abstract}
\end{equation}

While for the original QKD protocol the sifting ratio was 50\%, asymmetric implementations allow to improve  efficiency through biased basis selection. On the other hand, QBER describes the inconsistencies between the sifted bit streams held by the two parties, i.e.,

\begin{equation}
      E=\text{QBER} \triangleq \frac{\text{number of errors}}{\text{size of sifted key}}
      \label{eq:qber_or}
\end{equation}

After mismatches in measurement bases are discarded during sifting, errors in the two bitstreams are attributed to noise introduced due to channel imperfections (e.g., depolarization) or eavesdropping. QBER determines the fraction of the raw/sifted key that can be distilled to form the secret key by estimating i) the number of bits which need to be discarded during error correction, and ii) the amount of information which has potentially leaked to an adversary. Evidently, SKR is dependent on both raw key rate and QBER, as 

\begin{equation}
    \text{SKR}=R_{\text{raw}}\cdot \eta_{\text{sift}}\cdot g(E) 
\end{equation}

The analytical forms of $R_{\text{raw}}$ and $g$ depend on the specific QKD protocol and post-processing algorithms implemented. In a qubit-based system, the two terms are theoretically given as \cite{yuan_10-mbs_2018}:

\begin{equation}
    R_{\text{raw}}=f_{\text{source}}\cdot \eta_{d} \cdot t_{\text{chan}}
\end{equation}

\begin{equation}
    g(E, \bar{e_1}, \underline{p_1} )=\underline{p_1}\left[1-h\left(\overline{e_1}\right)\right]-f_{\text{ec}} h(E)
\end{equation}

Where $f_{\text{ec}}$ is the error correction efficiency, $\underline{p_1}$ is the lower bound for the probability that a sifted bit was registered from the detection of a single photon state, $\bar{e_1}$ is the upper bound for error rate in single photon states, and $E$ is the QBER from \autoref{eq:qber_abstract}. The detection efficiency is represented as $\eta_{d}$ and the channel transmissivity is given as:

\begin{equation}
   t_{\text{chan}}=10^{-al/10}
\end{equation}

\begin{figure}[h!]
    \centering
    \includegraphics[width=0.5\linewidth]{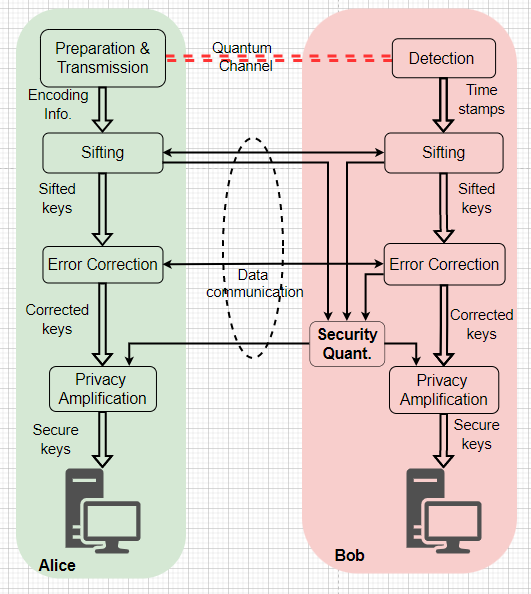}
    \caption{QKD key distillation procedure as presented in \cite{yuan_10-mbs_2018}.}
    \label{fig:qkd_loop_from_yuan_2018}
\end{figure}

Here, $l$ is the transmission distance and $a$ is the channel attenuation coefficient in dB per unit length. A milestone in practical QKD realization was the development of the decoy state technique \cite{hwang_quantum_2003, wang_beating_2005, lo_decoy_2005}. In decoy state protocols, signal photon states are randomly blended with the so-called decoy states to achieve higher key rates and security against advanced eavesdropping strategies, such as the Photon-Number Splitting (PNS) attack \cite{scarani_security_2009}. Therefore, an important contribution of the decoy state technique is the ability to approach the security of ideal QKD protocols while using practical, imperfect hardware. A schematic of the QKD key distillation procedure is shown in \autoref{fig:qkd_loop_from_yuan_2018}.

\section{Related work}
\label{sec:related_work}

Cybersecurity for the nuclear power sector is a field gaining interest, especially since the release of U.S. Nuclear Regulatory Commission (U.S. NRC) Regulatory Guide 5.71 in 2010 \cite{nrc_regulatory_2010}.  The guidance introduced a compliance-based security approach, defining critical systems and corresponding Critical Digital Assets (CDAs). The systems are classified in security levels, with restrictions applied to inter-level communication. A similar level-based security approach is proposed by the International Atomic Energy Agency (IAEA) \cite{agency_computer_2021}. 

Research has been conducted for vulnerability assessment \cite{peterson_overview_2019, song_analysis_2013} while intrusion detection systems have been proposed for software \cite{zhang_a_robust_2020, gkouliaras_false_2024} as well as hardware implementations \cite{roh_cyber_2020, allison_goosewolf_2023, karch_field_2024}. Several research works have explored classical encryption capabilities for nuclear power reactors. Emphasis has been given on hardware implementations, as with Field Programmable Gate Arrays (FPGAs) \cite{elakrat_development_2018,kumar_smart_2021} and Programmable Logic Controllers (PLCs) \cite{son_modeling_2023}.

QKD has received significant attention for critical infrastructure and energy sector \cite{paudel_quantum_2023}. Proposed application areas include the smart grid \cite{li_effect_2014, alshowkan_authentication_2022, grice_quantum_2025}, electrical utilities \cite{evans_trusted_2021}, and hydropower plants \cite{green_quantum_2023}. A detailed review of QKD energy applications is found in \cite{paudel_quantum_2023}. However, examining QKD potential for the nuclear power sector is a particularly underexplored topic, with related works limited in a preliminary concept study \cite{pantopoulou_monitoring_2024} and an investigation of wireless QKD application \cite{roberts_investigating_2021}. Our previous work included the development of a QKD simulator \cite{gkouliaras_nuqkd_2024}, leveraged to evaluate QKD performance for different nuclear environment use cases \cite{gkouliaras_exploring_2024}. 

As of yet, there has been no experimental demonstration of quantum nuclear reactor communications, to the best of our knowledge. While the nuclear industry shares characteristics with other energy sectors, it also exhibits certain distinguishing differences. Such traits justify the need to specifically evaluate QKD system in nuclear settings. Firstly, nuclear operation prioritizes plant availability, a principle that any non-safety systems (e.g., communication modules) need to comply with. In addition, nuclear systems have zero tolerance for data discontinuities, thus emphasizing the importance of timely, real-time operation. Finally, it remains to be shown whether radiation environments would affect QKD performance and whether such interference would be depicted in system metrics. The above arguments further highlight the potential impact of the present experiment.

\section{Secure communication model}
\label{sec:communication_model}

The goal of this work is to investigate the compatibility of QKD with nuclear reactor control systems, for realizing secure data exchange. The distinctiveness of such systems lies in two main characteristics, potentially differentiating them from other sector counterparts. Firstly, there is a requirement for 100\% system availability. Therefore, disrupting system operation is not a viable option except for scheduled and absolutely necessary tasks (e.g., refueling). Secondly, no data inconsistencies can be allowed, as the data historian needs to precisely archive past reactor states and be easily accessible for future reference.

To evaluate system performance under different use cases, we formulate a communication model tailored to nuclear reactor operations. The model defines a set of eight parameters with respect to data generation, encryption, and transmission. In addition, it describes the different stages of the data exchange loop, and mathematically derives the constraints for sustaining uninterrupted secure communication. As a result, it provides the tools to investigate the compatibility of a reactor-agnostic remote operation use case with an arbitrary QKD system.

\begin{table}[h!]
    \centering
    \caption{Summary of secure communication use-case parameters. }
    \begin{tabular}{|>{\centering\arraybackslash}m{2cm}|>{\centering\arraybackslash}m{3cm}|>{\centering\arraybackslash}m{2cm}|>{\centering\arraybackslash}m{6cm}|}
    \hline
    \textbf{Symbol} & \textbf{Name} & \textbf{Units}      & \textbf{Description} \\ %\hline

    \hline \hline
    \multicolumn{4}{|c|}{Data parameters} \\ \hline %\hline
    $N$             & Number of signals & Unitless      & Number of signals to be transmitted \\ \hline
    $f_s$           & Sampling rate & Cycles/s                 & Rate of controller signal update \\ \hline
    $f_{\text{rep}}$& Reporting rate& Cycles/s                 & Rate at which signals are sent to remote location \\ \hline
    $p$             & Precision     & Bits/value         & Number of bits allocated to each value \\ \hline \hline

    \multicolumn{4}{|c|}{Security parameters} \\ \hline %\hline
    
    $f_{\text{enc}}$& Key reusability factor & Unitless & Disposable key bits required per bit of encrypted data \\ \hline

    $t_{\text{auto}}$ & Communication autonomy & Minutes & Minimum secure operation time after key generation failure \\ \hline \hline

   \multicolumn{4}{|c|}{Channel parameters} \\ \hline %\hline

    $l$ & Channel length & Kilometers & Distance between two secure locations \\ \hline

    $E$ & Quantum Bit Error Rate & $\%$ & Noise level in the quantum channel \\ \hline

    \end{tabular}
    \label{tab:use_case_config}
\end{table}

\subsection{Communication parameters}

Model parameters are classified into three groups: data, security and channel. Data parameters are associated with the signal resolution and quantity, and they are determined based on domain knowledge and use case analysis. Security parameters are determined based on the use-case confidentiality and availability requirements. Finally, channel parameters are related to the properties of the QKD transmission link. The parameters are summarized in \autoref{tab:use_case_config}.

 The \emph{number of signals} ($N$) is determined by the amount of information a particular use case requires. While data historian requires the entirety of generated signals to be transmitted, a remote monitoring application might limit such number to include only a subset of signals necessary to verify normal operation.

The \emph{data sampling rate} ($f_s$) describes the frequency with which the reactor controller records field sensor data. Meanwhile, \emph{data reporting rate} ($f_{\text{rep}}$) is a quantity referring to the rate at which data are reported to the remote terminal unit. The two parameters do not need to be identical by default. For example, data could be sampled every 1 millisecond but reported once every 100 milliseconds. In this scenario, the remaining values are discarded. However, this might not always be the case; for instance, it could be possible that even though data are updated 10 times per second ($f_s=10$ Hz), they are reported once every second as a batch.

The \emph{precision} is the number of bits required to encode each data point in digital format. Precision is dependent on the data representation scheme, making assumptions on the floating point accuracy. Using the IEEE-754 standard for floating point arithmetic \cite{noauthor_ieee_2019}, the main options are either single precision (32 bits per value) or double precision (64 bits per value). 

Finally, \emph{key reusability factor} $f_{\text{enc}}$ describes the number of disposable key bits required to encrypt one bit of data, that is,

\begin{equation}
    f_{\text{enc}}=\frac{\text{key size}}{\text{encrypted data size}}\in (0,1]
    \label{eq:key_reusability_abstract}.
\end{equation}

The reusability factor depends on the cryptographic algorithm implemented. Lower values of the ratio represent higher key economy, potentially at the cost of theoretic security. Information-theoretic security requires one disposable key bit per encrypted data bit, thus $f_{\text{enc}}=1$. This is the case with OTP. All remaining symmetric cryptographic algorithms practically exhibit $f_{\text{enc}}<1$, as key material is somehow reused. In those cases, coupling with QKD offers only computational security. However, security bounds can be improved by increasing the key refresh rate, a task which is challenging when using conventional key distribution methods (e.g., public cryptography) \cite{gidney_how_2021}. Therefore, a QKD implementation has the potential to actually upgrade communication confidentiality, not only in the ideal scenario of coupling with OTP, but even when combined with applied encryption algorithms. 

The \emph{channel length} $l$ describes the distance between the reactor site and the remote location. Along with the \emph{error rate} (channel noise), it determines the secret key generation rate, as larger distances introduce increased photon attenuation in the optical fiber. Finally, \emph{communication autonomy} $t_{\text{auto}}$ describes the target system uptime following a potential failure of the key generation system (e.g., channel interception). During this interval, the system is expected to maintain secure communication using a reserve of secret keys.

\begin{figure}[h!]
    \centering
    \includegraphics[width=0.6\linewidth]{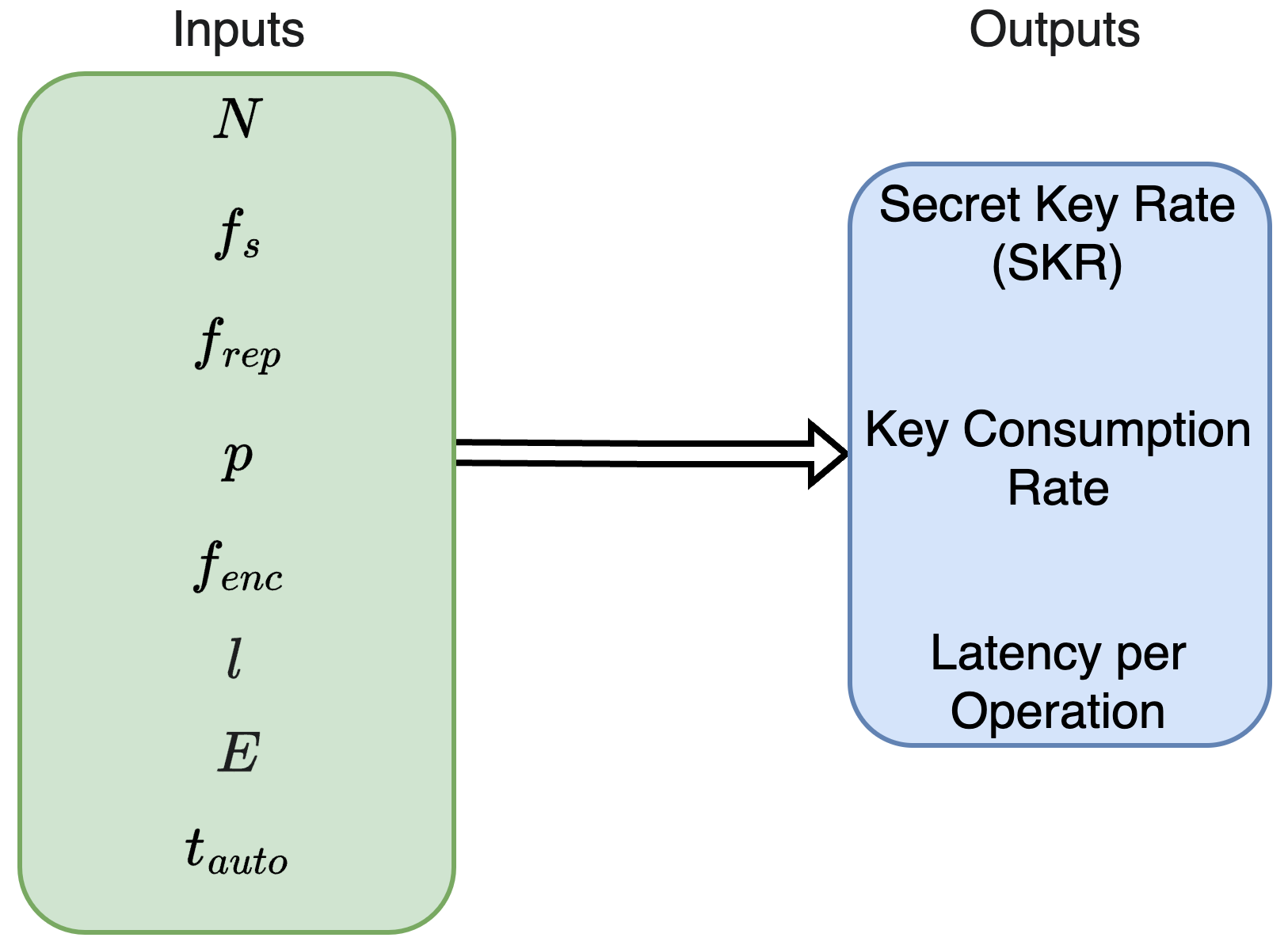}
    \caption{Communication model input and output parameters.}
    \label{fig:model_io}
\end{figure}

Based on the above parameters, a use case can be evaluated to determine the key consumption rate, the secret key generation rate, and the latency per operation. The model input and output parameters are shown in \autoref{fig:model_io}. As explained in the following sections, the outputs are evaluated based on several constraint conditions, to determine the feasibility of the use case.

\subsection{Procedure}

\begin{figure}[h!]
    \centering
    \includegraphics[width=0.8\linewidth]{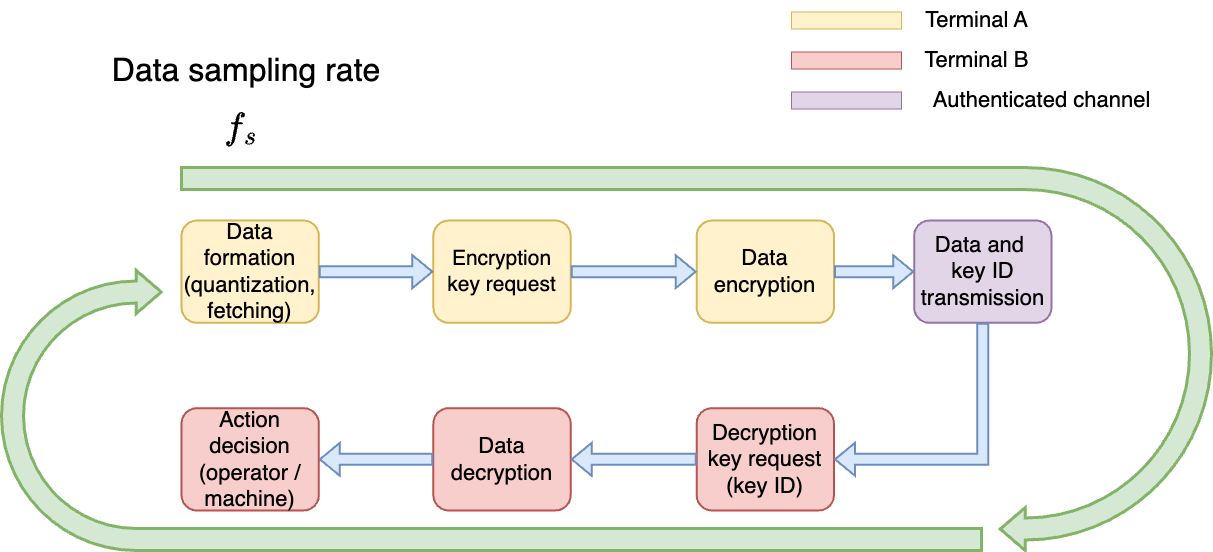}
    \caption{Data communication loop schematic. Terminal A is connected to the reactor PLC while Terminal B is the remote station. Total latency is the sum of latencies from each individual data processing stage. The target latency is determined by the data sampling rate $f_s$.}
    \label{fig:data_com_loop}
\end{figure}

The communication procedure stages are demonstrated in \autoref{fig:data_com_loop}. Terminal A fetches data from the reactor digital controller and encodes them in a digital representation scheme (e.g., IEEE-754). Following, it communicates with Key Server Alice requesting an encryption key, the size of which is dictated by the encryption algorithm and/or data size. Key Server Alice replies by sending the key and the associated key ID. The data are encrypted, and the ciphertext is transmitted to Terminal B over the authenticated channel along with the key ID. Upon receipt, Terminal B requests the decryption key from Key Server Bob, providing the ID. The ciphertext is decrypted and reactor data are extracted. Finally, an automated model (ML/AI, rule-based) analyzes the received data and provides some form of feedback to the operator regarding action to be taken. For example, the model might provide guidance for rod movement to match the target reactor power or neutron flux. %fission rate.

Two main metrics are used to evaluate system performance, latency and key availability. The parameters associated with each use case determine the target latency and target key availability. For a particular configuration to be realizable, \emph{both} latency and key availability conditions need to be satisfied. A third condition is related to meeting the target communication autonomy, in case of key distribution failure. Such conditions are mathematically derived in the following sections.

\subsection{Latency condition}

Latency is the first performance metric for evaluating the feasibility of a use case. The data sampling period ($\Delta t=1/f_s$) determines the time limit during which all stages need to have been completed, as reactor data are updated. In the scenario where data are transmitted to the remote terminal in fixed time intervals as batches, the threshold value is defined by the reporting period $1/f_{\text{rep}}=\kappa\cdot \Delta t$, where $\kappa \in \mathbb{N}^*$. Any excess time contributes to latency. As a result, the target latency condition can be expressed as:

\begin{equation}
t_{\text{fetch, A}}+t_{\text{key, A}}+t_{\text{enc, A}}+t_{\text{transm}}+t_{\text{key, B}}+t_{\text{dec, B}}+t_{\text{action, B}}\leq\text{max}(\frac{1}{f_s}, \frac{1}{f_{\text{rep}}})
\label{eq:latency_condition_original}
\end{equation}

Where each variable represents the elapsed time associated with each communication stage. Indexes A, B represent the terminal where the particular process occurs. Neglecting the time required for data analysis, \autoref{eq:latency_condition_original} can be written as:

\begin{equation}
    t_{\text{total}}=t_{\text{qkd}}+t_{\text{crypto}}+t_{\text{com}}\leq\Delta \tau_{\text{ef}}
    \label{eq:latency_condition_final}
\end{equation}

Where $t_\text{{qkd}}=t_\text{{key, A}}+t_\text{{key, B}}$ is the total time involving key request and delivery via QKD server communication, $t_\text{{crypto}}=t_\text{{enc}}+t_\text{{dec}}$ is the encryption/decryption interval, and $t_\text{{com}}=t_\text{{fetch,A}}+t_\text{{transm}}$ is the overall time required for data exchange (fetching from PLC and transmitting between Terminals A and B). The effective period $\Delta \tau_{\text{ef}}$ is defined as:   

\begin{equation}
    \Delta \tau_{\text{ef}}=\max\left(\frac{1}{f_s}, \frac{1}{f_{\text{rep}}}\right)
\end{equation}

The overall latency is dependent on the communication parameters (\autoref{tab:use_case_config}). %More detailed expressions can be derived for individual use cases.

\subsection{Key availability condition}

The target key availability condition dictates that the secret key formed during the data generation interval should be at least equal in size to the key $n$ required to encrypt the data. Therefore, we can write:

\begin{equation}
    n(N, p, f_{\text{enc}})\leq \text{SKR}(l, E, t) \cdot \Delta \tau_{\text{ef}}
    \label{eq:bandwidth_condition_raw}
\end{equation}

Where $l$ and $E$ are the distance and error rate of the quantum channel, respectively. SKR is the corresponding secret key rate reported in the effective period $\Delta \tau_{\text{ef}}$, and is dependent on the channel parameters (length, error rate). The required key size $n$ is a function of the data and security parameters. For each reporting period, the amount of generated data in bits is: 

\begin{equation}
n_{\Delta \tau_{\text{ef}}}=N \cdot p \cdot \frac{f_s}{f_{\text{rep}}}=N \cdot p \cdot  f_s \cdot \Delta \tau_{\text{ef}} = N_{\text{ef}}\cdot p 
\label{eq:data_size}
\end{equation}

Where the effective number of signals is defined as:

\begin{equation}
N_{\text{ef}}=\kappa \cdot N
\end{equation}

The number of key bits needed per effective period is therefore:

\begin{equation}
n=n_{\Delta \tau_{ef}} \cdot f_{\text{enc}}
\label{eq:data_size_w_fenc}
\end{equation}

Where the key reusability factor $f_{\text{enc}}$ was given in \autoref{eq:key_reusability_abstract}. The condition of \autoref{eq:bandwidth_condition_raw} constitutes the tightest bound, assuming that any part of the key not consumed during the data reporting period is discarded. In practice, excess keys will be stored in the key management system contributing to a key reserve pool. Therefore, a more realistic condition is to ensure that the dynamic key pool size remains positive during all times. 

The size of the dynamic key pool $d$ can be thought of as the difference between key material contributed (from QKD) and key material consumed (from the secure application). Generated keys are added to the key pool in irregular times $t_g$ (once the key distillation iteration is completed and the secret key is formed), where $g\in\{0,1,2,...\}$ is the QKD cycle index. The key generation intervals are thus defined as:

\begin{equation}
    \Delta t_g =t_{g+1}-t_g
\end{equation}

Furthermore, key requests also occur periodically according to the data sampling/reporting rate. The duration of key generation intervals typically exceeds the nuclear data reporting period ($\Delta t_g>> \Delta \tau_{\text{ef}}$). As a result, the dynamic key pool size is given as:

\begin{equation}
    d[k]=d[0]+ \sum_{g\in G(k)} (\Delta t_g \cdot \text{SKR}_g) - \sum_{i=0}^{k}n[i],  \quad \forall k\in\{0,1,2,...\}
    \label{eq:sk_balance}
\end{equation}

 Where $k=t/\Delta \tau_{\text{ef}}$ is the communication loop index with respect to \autoref{fig:data_com_loop} and:

\begin{equation}
G(k) = \{ g : t_{g+1} \leq k \cdot \Delta \tau_{\text{ef}} \}
\end{equation}

 \autoref{eq:sk_balance} assumes that key consumption $n$ could vary over time, due to a potential change in communication parameters. Here, the time step $k=0$ marks the initiation of reactor operation and data exchange. The initial pool size $d[0]$ is related to the QKD operational time before the beginning of data encryption (lead time) or the size of pre-shared symmetric keys. Therefore, the updated target key availability  condition is written as:

\begin{equation}
    d[k]>0 , \quad \forall k\in\{0,1,2,...\}
    \label{eq:target_bandwidth}
\end{equation}

The dependence of the key availability condition on previous time terms guarantees that the encryption module treats newly generated data in chronological order. Thus, evaluating \autoref{eq:target_bandwidth} for a given time step $k$ satisfies that all data generated since the beginning of operations have been successfully processed.

\subsection{Lead time and post-failure uptime}

As availability remains one of the main priorities in critical infrastructure, particularly in the nuclear power sector, any remote operation needs to be designed in a manner minimizing the probability of reactor shut down. Due to the fact that fission products are strong neutron absorbers (e.g., Xenon-135), the reactor cannot be instantly restarted (reactor poisoning).  Therefore, an emergency shutdown could potentially have broad implications by failing to meet the energy demand for a prolonged time period. From a cybersecurity perspective, it becomes clear that shutting the plant is not an acceptable cyber event response strategy, and should only take place if absolutely necessary.  To maintain uninterrupted secure communication, two related parameters are defined, the QKD lead time and post-failure operational time. 

Lead time describes the interval before the initiation of secure communication, during which key distribution is operating. During the lead time interval, QKD contributes secret keys to form the reserve key pool, i.e.,:

\begin{equation}
d_0=d[k_{\text{lead}}]=\sum_{g\in G(k_{{lead}})} (\Delta t_g \cdot \text{SKR}_g) %=\sum_{i=0}^{k_{\text{lead}}}SKR[i]
\label{eq:sk0}
\end{equation}

Where $k_{\text{lead}}=t_{\text{lead}}/\Delta \tau _{\text{ef}} $. To explicitly account for the QKD lead time, the dynamic pool size of \autoref{eq:sk_balance} can be expressed as:

\begin{equation}
    d[k]=\sum_{g\in G(k)} (\Delta t_g \cdot \text{SKR}_g) - \sum_{i=k_{\text{lead}}+1}^{k}n[i],  \quad \forall i\in\{0,1,...,k\}
    \label{eq:sk_balance_w_lead_time}
\end{equation}

The initial key reserve can be approximated based on the time-averaged SKR and lead time as:

\begin{equation}
    d_0 \approx \overline{\text{SKR}}(l, E)\cdot t_{\text{lead}}
    \label{eq:t_lead}
\end{equation}

The goal of the initial reserve is to provide a safety margin for balancing key consumption requirements, during the initial stages of secure communication. As contributions to the key pool take place at discrete times, meeting the key availability condition might be challenging even if the average SKR is higher than the key consumption rate. The assignment of a QKD lead time helps maintain the positive size of the key pool throughout the reactor operation. Based on this criterion, \autoref{sec:qkd_lead} determines the minimum lead time for a variety of parameter combinations.

A second benefit of the initial reserve is prolonging secure communication in case of an emergency failure of the key distribution system. We refer to the elapsed time between QKD failure and the exhaustion of key material as post-failure uptime, defined as: 

\begin{equation}
\Delta t_{\text{up}}=t_{\text{d=0}}-t_\text{{fail}}
\end{equation}

Here, $t_{\text{d=0}}$ is the time when the key reserve reaches zero and $t_\text{{fail}}$ is the QKD failure instance. $\Delta t_{\text{up}}$ is a function of the secure communication parameters, including the lead time. In \autoref{sec:post_failure}, post-failure uptimes are evaluated for various parameter configurations.

A third condition can be formulated by defining a minimum acceptable communication autonomy (threshold uptime) $t_{\text{auto}}$, such that:

\begin{equation}
    \Delta t_{\text{up}} \geq t_{\text{auto}}
        \label{eq:secure_uptime}
\end{equation}

\autoref{eq:secure_uptime} constitutes the secure uptime condition. With respect to the key pool size at failure, it can be equivalently written as:

\begin{equation}
    d[k_{\text{fail}}] \geq \frac{n_{\text{pf}} \cdot t_{\text{auto}}}{\Delta \tau _{\text{ef}}}
    \label{eq:pf_condition}
\end{equation}

Where $k_{\text{fail}}=t_{\text{fail}} / {\Delta \tau _{\text{ef}}}$ is the discrete time index of the key distribution failure instance. $n_{\text{pf}}$ is the post-failure key consumption rate per effective period. Although $n_{\text{pf}}$ could be equal to the pre-failure key consumption rate, this might not always be the case; to maximize the uptime, it is possible that during the outage only a subset of critical signals are transmitted. Similarly, the encryption algorithm could be switched to a practical cipher to achieve higher key economy. For convenience, we define the generalized key consumption rate as:

\begin{equation}
\Tilde{n}[k]
\begin{cases} 
0, & \text{for } 0<k<k_{\text{lead}} \\
n, & \text{for } k_{\text{lead}} \leq k \leq k_{\text{fail}} \\
n_{\text{pf}}, & \text{for } k > k_{\text{fail}}
\end{cases}
\label{eq:generalized_consumption}
\end{equation}

Substituting \autoref{eq:sk_balance_w_lead_time} and \autoref{eq:generalized_consumption} into \autoref{eq:pf_condition}, the final inequality becomes:

\begin{equation}
    \sum_{g\in G(k_{\text{fail}})} (\Delta t_g \cdot \text{SKR}_g) - \sum_{i=0}^{k_{\text{fail}}}\Tilde{n}[i] \geq \sum_{i=k_{\text{fail}}+1}^{k_{\text{auto}}} \Tilde{n}[i]
    \label{eq:pf_condition_final1}
\end{equation}

Where $k_{\text{auto}}=t_{\text{auto}}/\Delta \tau_{\text{ef}}$. The above inequality thus evaluates both the key availability and secure uptime conditions. Grouping the $\Tilde{n}$ terms, we obtain:

\begin{equation}
    \sum_{g\in G(k_{\text{fail}})} (\Delta t_g \cdot \text{SKR}_g) - \sum_{i=0}^{k_{\text{auto}}}\Tilde{n}[i] \geq 0
    \label{eq:pf_condition_final2}
\end{equation}

To obtain a practical estimate when a time-dependent dataset is not available, the sum terms can be replaced by the average values to obtain the simplified expression of \autoref{eq:pf_condition_simplified}:

\begin{equation}
 t_{\text{fail}} \left[\overline{\text{SKR}}(l, E) - \frac{\bar{n}}{\Delta \tau _{\text{ef}}}\left(1 - \frac{t_{\text{lead}}}{t_{\text{fail}}} \right) \right] \geq \frac{\bar{n}_{\text{pf}} \cdot t_{\text{auto}}}{\Delta \tau _{\text{ef}}}
\label{eq:pf_condition_simplified}
\end{equation}

Equations \ref{eq:pf_condition_final1}- \ref{eq:pf_condition_simplified} can be handful to determine the minimum QKD lead time required to achieve a specific post-failure uptime, and vice versa. 

In summary, for a use case to be feasible, the model outputs need to satisfy the latency condition (\autoref{eq:latency_condition_final}) and the key availability condition (\autoref{eq:bandwidth_condition_raw} or \autoref{eq:target_bandwidth}). Given the fulfillment of the two constraints, a minimum operational lead time can be determined from  \autoref{eq:pf_condition_simplified} to satisfy the specified communication autonomy requirement and prolong system operation following a potential key distribution failure.

\section{Use cases}
\label{sec:pur1}

\begin{figure}[ht]
    \centering
    \includegraphics[width=0.7\linewidth]{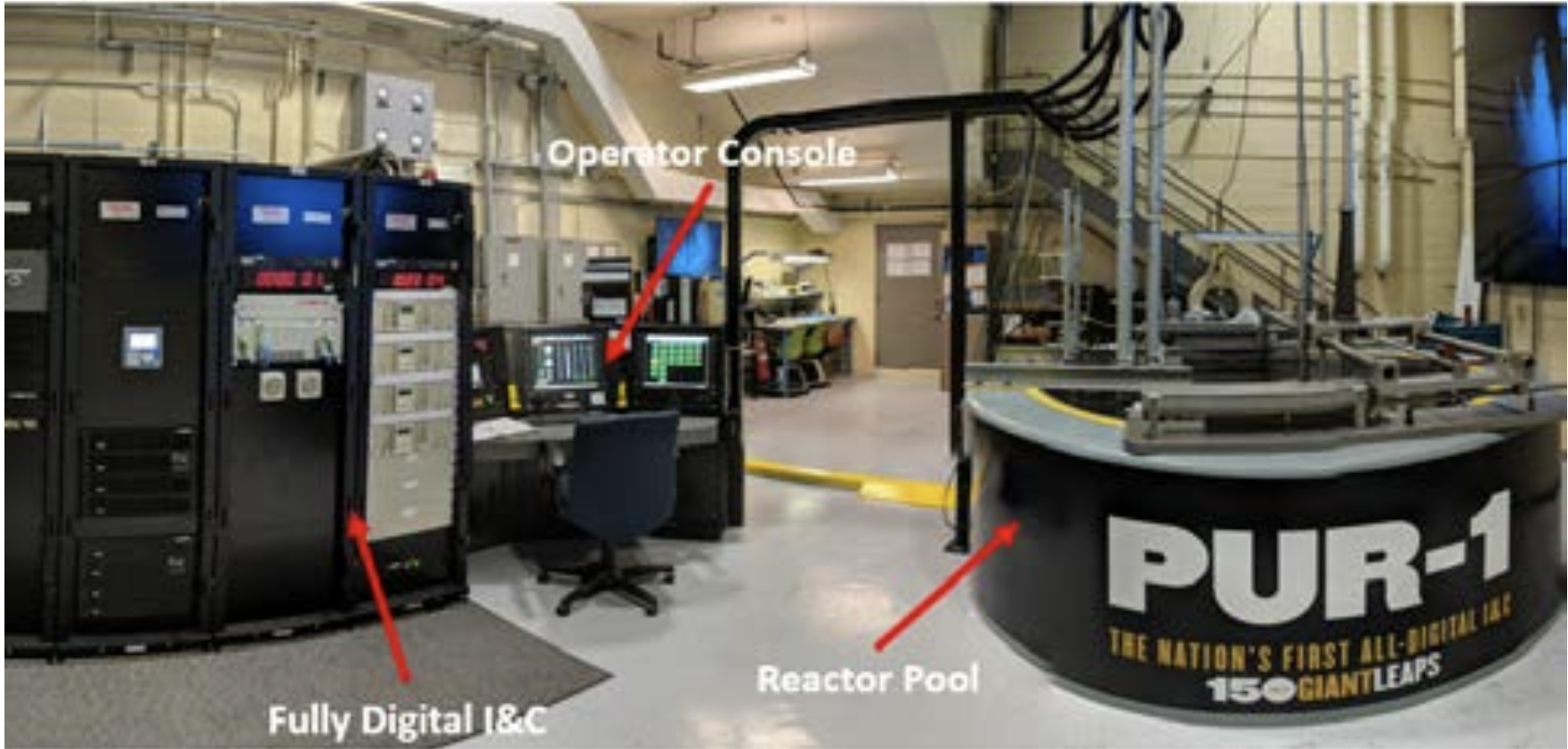}
    \caption{PUR-1 reactor room.}
    \label{fig:pur1_photo}
\end{figure}

PUR-1 is a pool-type research reactor located at Purdue University. To date, it is the only fully digital reactor licensed by the US NRC. As the future of the nuclear power sector is associated with digitalization, PUR-1 acts as a prototype of I\&C architectures to be incorporated in anticipated realizations of advanced reactor designs (microreactors, SMRs). Consequently, an implementation based on PUR-1 is ideal to gain maximum insight on future reactor challenges, from a control and monitoring perspective. The PUR-1 reactor room is shown in \autoref{fig:pur1_photo}.

The Programmable Logic Controller (PLC) allows remote monitoring and collection of more than 2,000 parameters, including digital values (e.g., manual SCRAM control) and digitized analog quantities (e.g., neutron flux). Of these, 67 signals have been found to be most relevant for representing important system behavior. This selection has been conducted based on domain knowledge, excluding parameters not directly related to the power generation process (e.g., room temperature) \cite{chatzidakis_characterizing_2024}. A timestamp is also included to form a set of 68 signals in total.

Based on this classification, two general categories of communication use cases can be defined, remote monitoring and data historian. In the first category, a remote operator needs to obtain real-time information on the reactor state to allow potential action to be taken. The main priority is to have minimum latency and zero delay, while only relevant signals are required. As a general approach, this use case benefits from a higher sampling rate to obtain adequate resolution in the time domain. In the second category, the complete set of signals needs to be transmitted to the remote server. On top of documentation purposes, the entirety of data would be used for providing long-term analytics (e.g., for load following) and for training Digital Twin models. Therefore, all signals generated per time step need to be remotely transmitted. Based on the specifics of the system, the sampling rate can be selected optimally to satisfy time resolution requirements, without consuming excess storage space in the long term. The two use cases are summarized in \autoref{tab:use_cases}.

\begin{table}[h!]
    \centering
    \caption{PUR-1 remote operation use cases.}
    \renewcommand{\arraystretch}{1.5} % Adjust row height
    \begin{tabular}{|>{\centering\arraybackslash}m{1cm}|>{\centering\arraybackslash}m{2cm}|>{\centering\arraybackslash}m{1.75cm}|>{\centering\arraybackslash}m{2cm}|>{\centering\arraybackslash}m{2cm}|>{\centering\arraybackslash}m{5cm}|}
        \hline 
        Use case & Name & Number of signals & Latency priority & Sampling rate & Description \\
        \hline
        \hline 
        \#1 & Remote monitoring & 68 & High & High & Core signals transmitted to remote operator. \\
        \hline 
        \#2 & Data historian & 2,000 & Medium & Medium & All signals transmitted for remote storage. \\
        \hline
    \end{tabular}
    \label{tab:use_cases}
\end{table}

The control system supports arbitrary sampling frequencies, limited by the sensor capabilities. Typical sampling frequencies are found in the range of less than ten samples per second. Regarding data resolution, it has been shown in \cite{gkouliaras_exploring_2024} that single precision (32 bits) is sufficient to quantize and digitally encode reactor-generated data.

\section{Experimental setup}
\label{sec:experimental}

The experimental setup leverages PUR-1 and the commercial Toshiba Long Distance QKD system (QKD-LD). QKD-LD consists of four devices in total: QKD-Alice, QKD-Bob, Key Server Alice and Key Server Bob. QKD-Alice and QKD-Bob connect using a quantum and a classical channel, both of which are single-mode optical fibers. For the quantum communication stage, QKD-LD implements the T12 QKD protocol \cite{lucamarini_efficient_2013,yuan_10-mbs_2018}. T12 is a modified version of phase-encoding decoy state BB84, using asymmetric basis selection. One decoy and a vacuum state are blended with signal states with probabilities of 1.661\% and 1.466\%.

\begin{figure}[h!]
    \centering
    % First subfigure
    \begin{subfigure}[b]{0.39\textwidth}
        \centering
        \includegraphics[width=\linewidth]{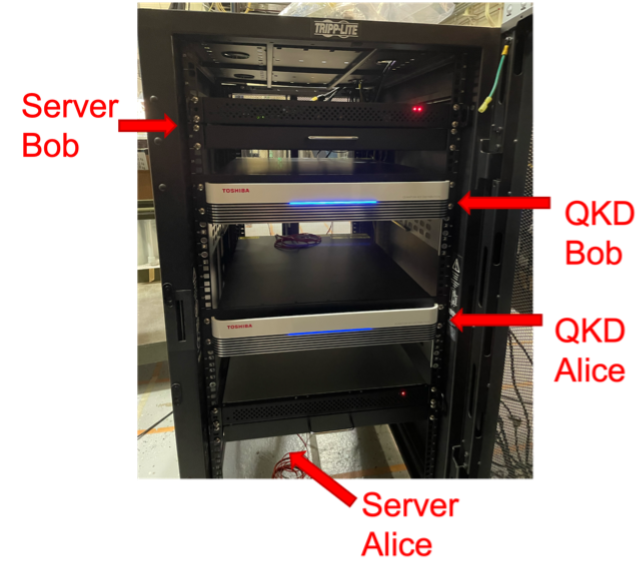}
        \caption{QKD stack.}
        \label{fig:qkd_reactor_1}
    \end{subfigure}
    \hfill % Adds horizontal spacing between subfigures
    % Second subfigure
    \begin{subfigure}[b]{0.6\linewidth}
        \centering
        \includegraphics[width=\textwidth]{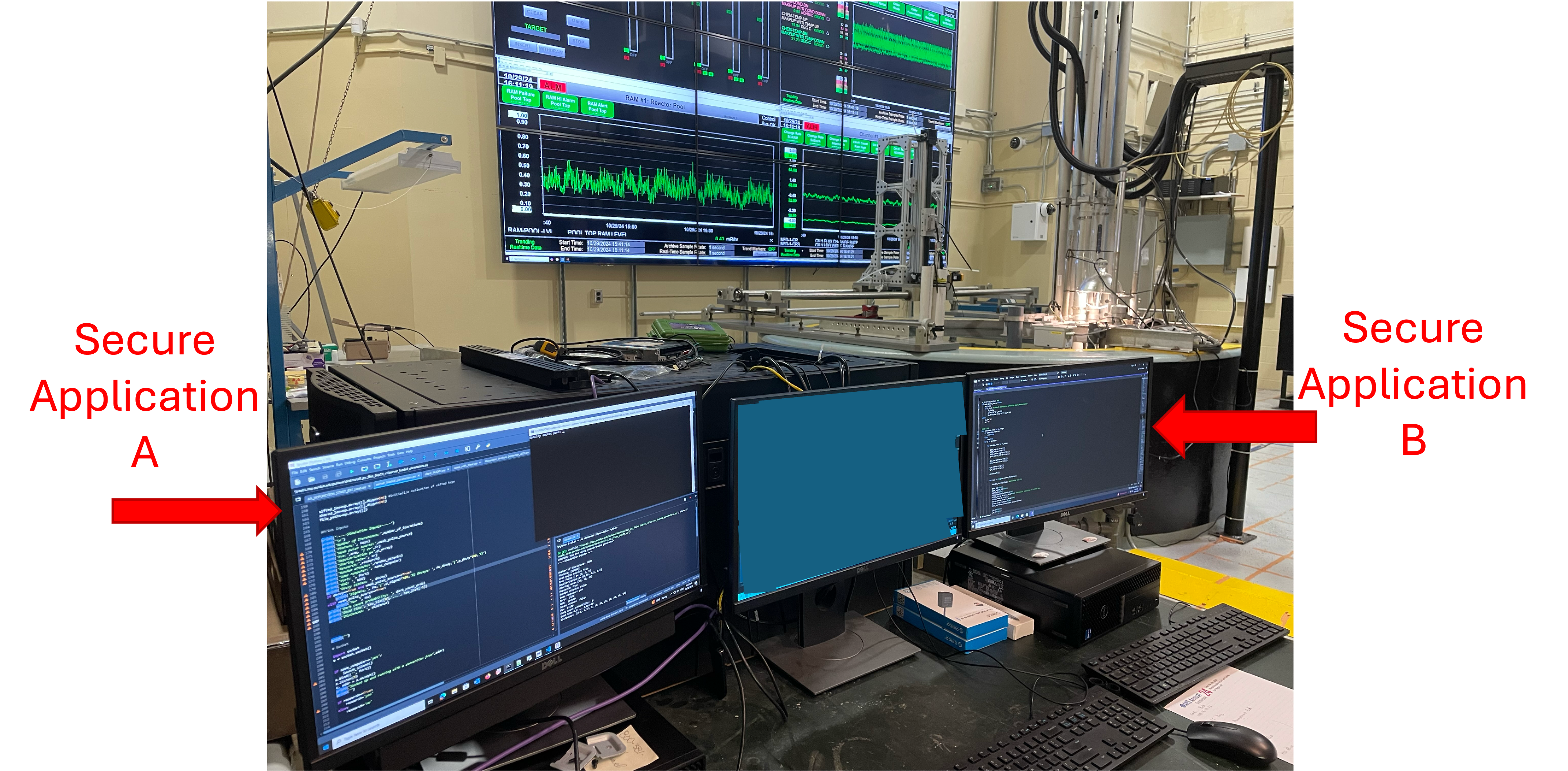}
        \caption{Encryption/Decryption workstations.}
        \label{fig:qkd_reactor_2}
    \end{subfigure}
     \hfill
    \begin{subfigure}[b]{0.49\linewidth}
    \centering
    \includegraphics[width=\textwidth]{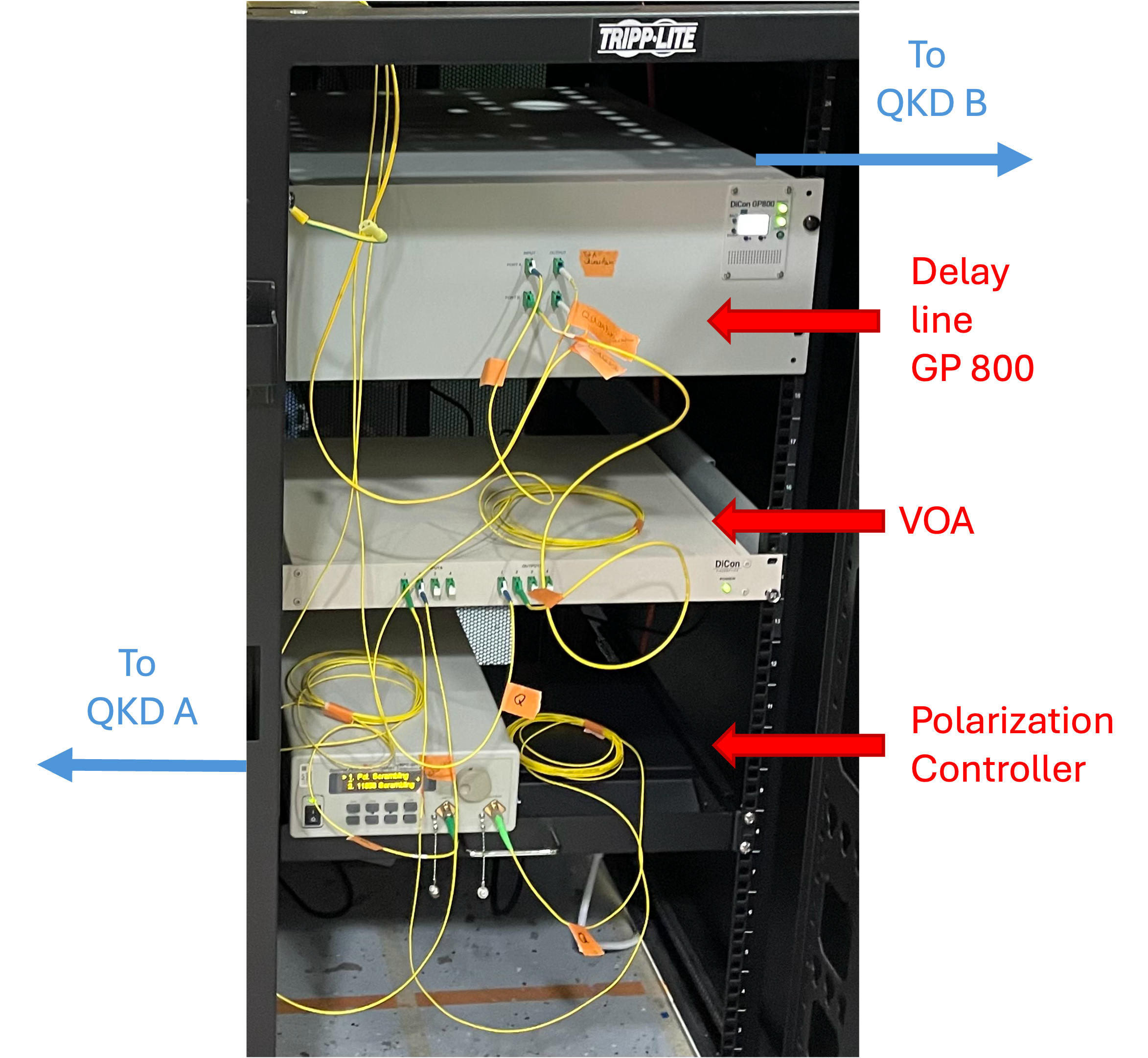}
    \caption{Equipment intercepting quantum and classical channel. Red arrows identify the devices, while blue arrows indicate integration with the QKD system.}
    \label{fig:ornl_stack}
\end{subfigure}

    \caption{QKD installation in PUR-1 control room.}
    \label{fig:qkd_reactor}
\end{figure}

\begin{figure}[h!]
    \centering
    \includegraphics[width=\linewidth]{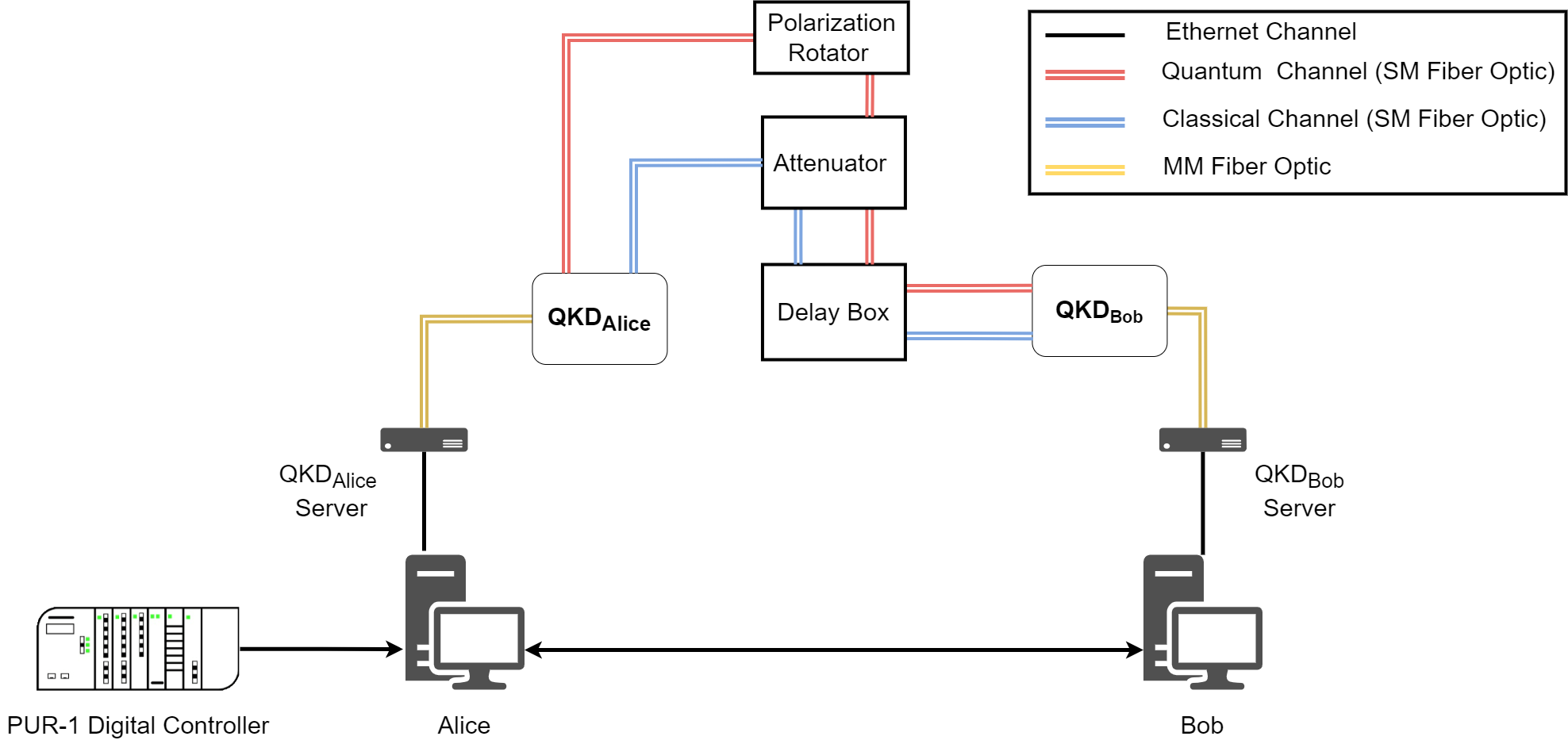}
    \caption{Schematic of PUR-1 QKD experiment setup. PUR-1 data are provided to sender (Alice) workstation (WA). WA requests keys from QKD server A to encrypt data. Encrypted data and key ID are transmitted over an authenticated channel. Receiver workstation (WB) requests the key from QKD server B by providing the key ID. The key is applied to received data for decryption according to the applied cryptographic scheme.}
    \label{fig:toshiba_schematic}
\end{figure}

\autoref{fig:qkd_reactor} displays the prototypic QKD setup installed in PUR-1 reactor, while \autoref{fig:toshiba_schematic} provides a schematic of the system. The system is structured around two workstations for the sender (Alice) and receiver (Bob), referred to as WA and WB. WA and WB represent the two remote locations, with WA physically connected to PUR-1, thus having access to reactor data. The two workstations run Windows 11 OS and communicate over a regular TCP/IP, non-dedicated LAN connection, shared with other network devices. Their purpose is to perform encryption and decryption operations, respectively, on real-time reactor data.

WA and WB are connected to the QKD servers of Alice and Bob, respectively, over dedicated Ethernet CAT-5e data links. The servers, running Linux OS, are responsible for implementing the Key Management System (KMS) by storing the QKD-generated keys and providing them upon request to the workstations, according to the ETSI GS QKD 014 standard \cite{noauthor_etsi_2019}. Whenever one of the two workstations requests a key, the corresponding KMS server replies with a unique key ID associated with the key. The key ID, a 128-bit Universally Unique Identifier (UUID), is simultaneously forwarded to the second server. The remote workstation receives the key ID over the authenticated channel, and uses it to request the corresponding key from the KMS server. Through this process, the two communicating parties gain access to an identical key to subsequently use with a symmetric encryption scheme. Since a single key cannot be requested twice and the communication between workstation and server is authenticated, the keyID does not need to be encrypted.

The two QKD devices (QKD-Alice, QKD-Bob) are connected to Key Server Alice and Key Server Bob, respectively, through duplex multi-mode fiber. Following the QKD protocol specification, they are connected to each other through a quantum and a classical channel. Both channels are single mode fibers (Corning SMF-28) terminated with LC/UPC connectors. The two channels should be identical in terms of length.

To experimentally replicate different transmission distances and environmental conditions, additional equipment intercepts both SMF channels. Three devices are connected sequentially, a General Photonics Polarization Controller, a DiCon Variable Optical Attenuator (VOA), and a DiCon GP800 delay line. \autoref{fig:ornl_stack} shows the stack of equipment intercepting the QKD channels.

The polarization controller enables manually controlling photon polarization by defining a voltage applied to a series of fiber squeezers. The VOA introduces channel attenuation up to 30 dB per channel, in order to replicate the effect of longer transmission distances. Additionally, two fixed 10 dB attenuators are introduced for the classical and quantum channel. Finally, the delay line uses MicroElectroMechanical (MEMS) optical switches to actually vary the fiber length connected, ranging from 0 to 32 km. The significance of the delay line stems from the fact that its effect is identical to adding different fiber segments, introducing not only photon attenuation but potential time/frequency dispersion effects. 

The interfering modules are operated under different parameter combinations to generate several channel use cases. Given that typical loss in SMF operating at a wavelength of 1550 nm is approximately 0.2 dB/km \cite{agrawal_fiber-optic_2010}, photon attenuation of a channel up to 200 km can be replicated. Combining the VOA and delay line allows for a more realistic fiber implementation while also further increasing the channel length capability. 

The TOSHIBA system reports the Secret Key Rate (SKR) and Quantum Bit Error Rate (QBER), along with the corresponding timestamp. QKD data are reported after each key distillation cycle is completed. Processing the data, it is possible to identify the random key material contributed to the pool as a function of time.

\section{Secret key generation}
\label{sec:secret_key_generation}

For the first stage of the experiment, QKD is operated at various channel lengths for prolonged periods of time. The distance was varied by modifying the VOA and delay line parameters. For each replicated fiber length, measurements were taken for a period of at least 10 hours of uninterrupted operation. 

\begin{figure}[h!]
    \centering
    \includegraphics[width=0.7\linewidth]{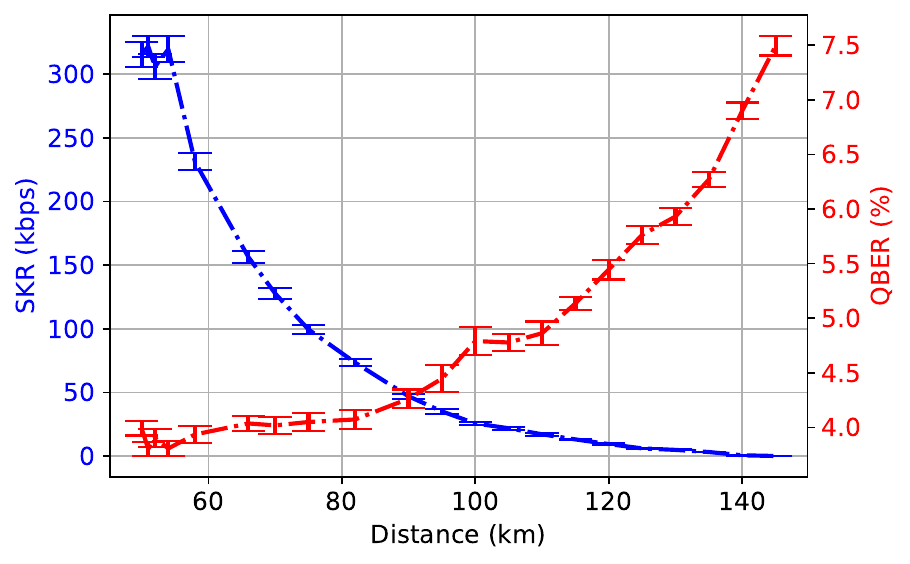}
    \caption{Average SKR and QBER as a function of distance. QKD data collected over 10 hours of operation at each length. Distances replicated using combinations of delay line and attenuators. Rate exceeds 315 kbps  at 50 km (10 dB loss). System does not generate secure keys at a distance of 145 km, which aligns with the 30 dB loss design of the Toshiba QKD LD system.  We observe that the error rate mostly increases with distance, reaching approximately 7.5\% at 145 km.}
    \label{fig:toshiba_skr_qber}
\end{figure}

\subsection{QKD performance evaluation}

\begin{figure}[h!]
    \centering
    \includegraphics[width=0.7\linewidth]{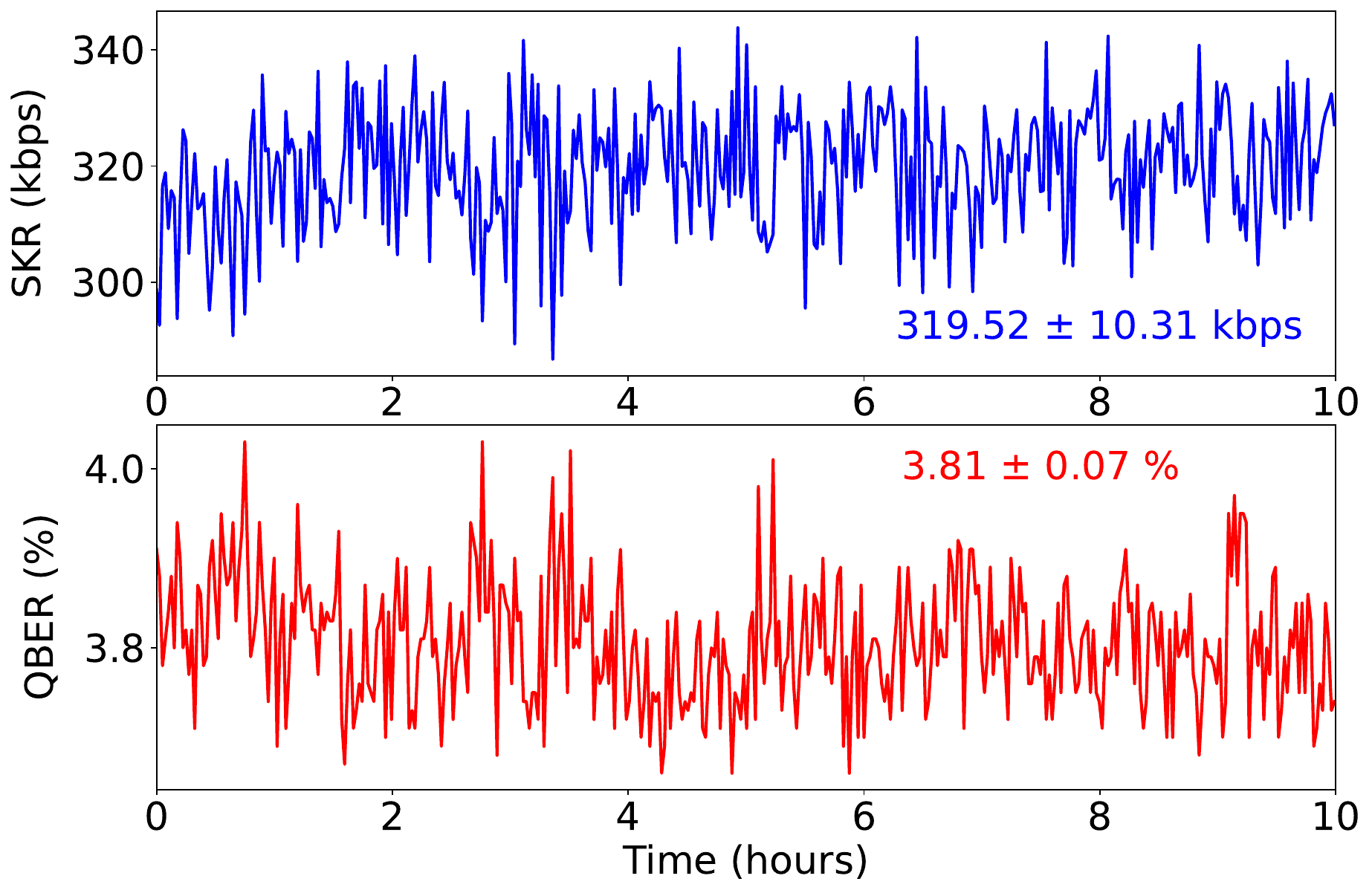}
    \caption{SKR and QBER versus time for 10 hours of operation ($l=54$ km). }
    \label{fig:toshiba_combined}
\end{figure}

SKR and QBER values are averaged over time and are displayed in \autoref{fig:toshiba_skr_qber}. An average rate of approximately 315 kbps is obtained at 50 km, declining exponentially as the channel length increases. At 145 km, no secret key can be distilled, leading to zero secret key rate. Performance aligns with the 30 dB loss design of the Toshiba QKD LD system. QBER ranges from approximately 4\% at 50 km to 7.5\% at 145 km. The curve demonstrates that, despite channel length and QBER not being directly related, increasing the distance contributes to higher error rate in practice. The total absence of key generation at 145 km is thus attributed not only to increased photon attenuation, but also to high error rate on the pulses actually detected, which prevents secure key distillation. Both SKR and QBER exhibit statistical fluctuations over time, as shown by the uncertainties in the plot. Indicatively, the two metrics at $l=54$ km are plotted versus time for the 10-hour interval, and shown in \autoref{fig:toshiba_combined}. The variance is approximately 10.3 kbps for SKR and only 0.07\% for the error rate.

\begin{figure}[ht]
    \centering
    \includegraphics[width=0.7\linewidth]{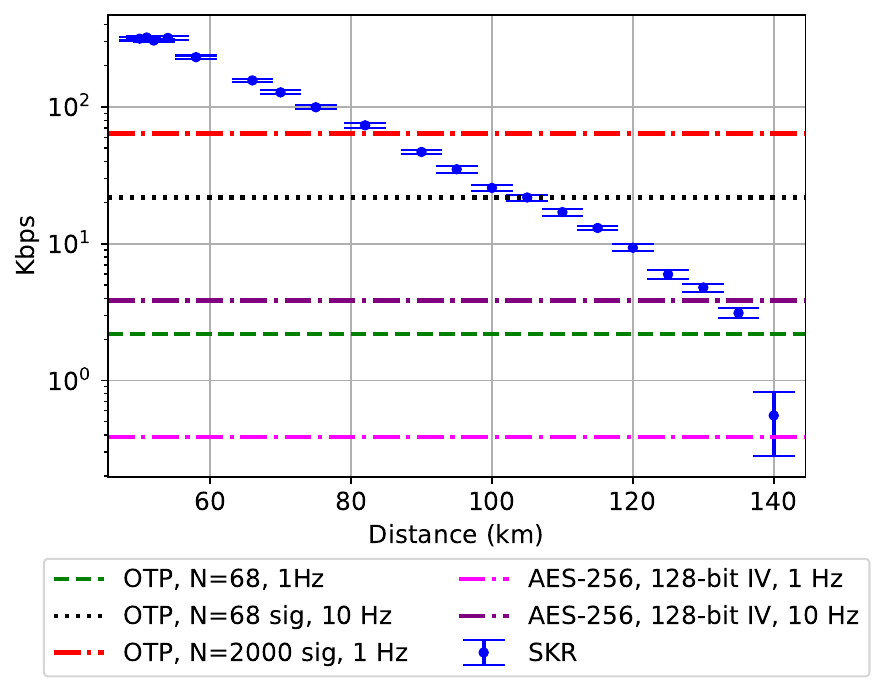}
    \caption{Maximum distance satisfying target key availability condition for indicative communication use cases. For $N=2,000$ signals and $f_s=f_{\text{rep}}=1 $ Hz, the tight bandwidth condition of \autoref{eq:bandwidth_condition_raw} is fulfilled at up to 82 km. Excess keys and initial key pool size are ignored. Distances up to 140 km can be achieved either by selecting a lower sampling rate or by switching to AES from OTP.}
    \label{fig:toshiba_skrs_time_use_cases}
\end{figure}

The goal is to determine the maximum distance satisfying the target key availability condition for different use cases.  The tight bound of \autoref{eq:bandwidth_condition_raw} is used for a first estimate, assuming zero initial key reserves and that excess keys are discarded. \autoref{fig:toshiba_skrs_time_use_cases} demonstrates how the recorded average SKRs compare to representative key consumption rates. Rates are calculated through \autoref{eq:data_size}, assuming OTP encryption ($f_{\text{enc}}=1$) with equal reporting and sampling frequencies ($f_s=f_{\text{rep}}$). Following this conservative approach, it is shown that the target key availability can be achieved for up to approximately 85 km and 2000 signals. If only the 68 core signals are transmitted, the maximum achievable distance is approximately 135 km and 105 km, for $f_s=1$ Hz and $f_s=10$ Hz, respectively. 

The plot also features two AES-256 configurations, at $f_s=1$ Hz and $f_s=10$ Hz. The key (256 bits) and the initialization vector (IV, 128 bits) are constantly updated, at the same rate as data sampling. Since the key/IV size is fixed per encrypted block, the target bandwidth remains constant for any number of signals, and is only dependent on the sampling/reporting rate. Even though AES is not information-theoretically secure, it is considered a -computationally speaking- robust encryption standard, officially adopted by NIST \cite{national_institute_of_standards_and_technology_us_advanced_2023}. As demonstrated in the plot, the application of AES-256 allows to extend the distance up to 140 km for an arbitrary number of signals, without lowering the sampling frequency.

\subsection{Dynamic key pool size}

\begin{figure}[ht]
    \centering
    \includegraphics[width=0.55\linewidth]{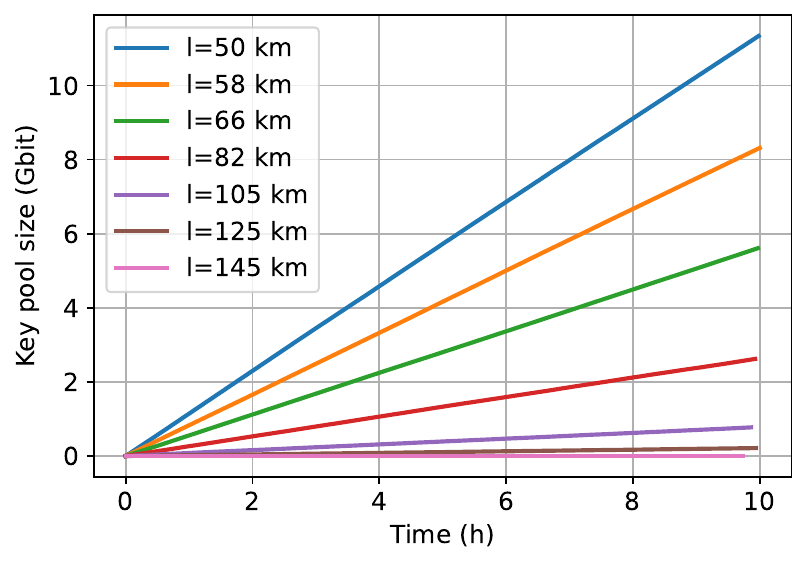}
    \caption{Accumulation of generated keys as a function of time for various distances.}
    \label{fig:static_key_pool}
\end{figure}

The secure communication system of \autoref{fig:toshiba_schematic} can be thought of as a combination of two interconnected but independent systems, the key generation system and the key distribution system. The key generation system, featuring the quantum layer, is responsible for creating a symmetric key reserve pool and continuously contributing the newly formed keys to it. The key pool is identical on the sides of Alice and Bob, stored in both servers. The key distribution system is responsible for removing keys from the pool upon request from the communication parties and encryption applications. The size and number of removed keys, as well as the frequency at which this process occurs, are dictated by the communication model parameters (\autoref{tab:use_case_config}). The system can be described by the balance \autoref{eq:sk_balance_w_lead_time}. We begin by studying the dynamics of the key pool size when isolated from the key distribution system (i.e., before secure communication/encryption is initiated). Processing the SKR data from the system, the curves of \autoref{fig:static_key_pool} are obtained.

\begin{figure}[h]
\centering
    \begin{subfigure}{0.49\linewidth}
        \centering
        \includegraphics[width=\linewidth]{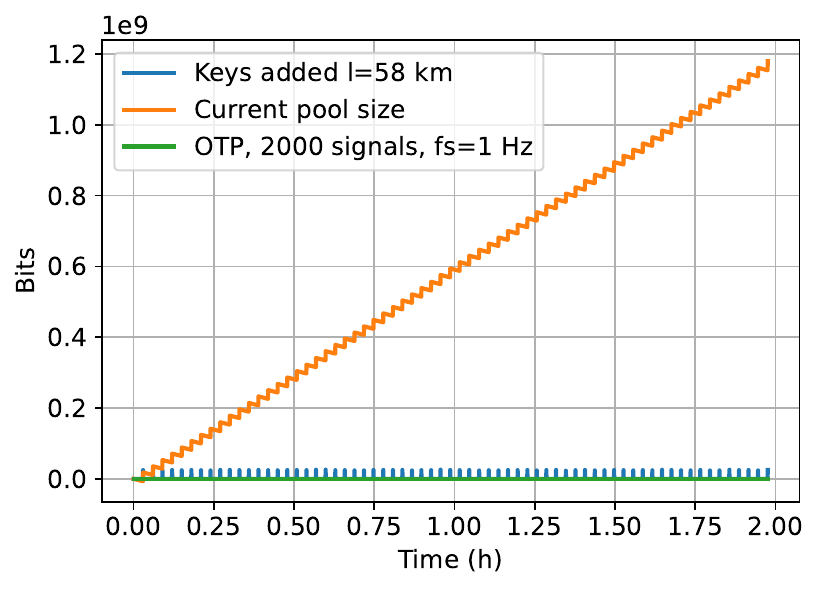} 
        \caption{$l=58$ km}
        \label{fig:subfig3}
    \end{subfigure}
    \hfill
    \begin{subfigure}{0.49\linewidth}
         \centering
        \includegraphics[width=\linewidth]{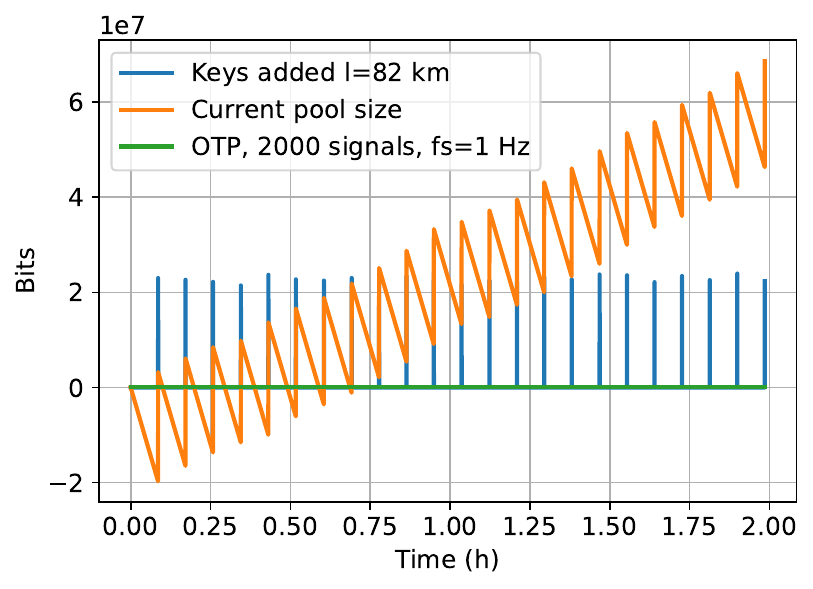} 
        \caption{$l=82$ km}
        \label{fig:subfig6}
    \end{subfigure}

    \caption{Dynamic key pool size for different communication distances. The practical-dynamic key availability condition of \autoref{eq:target_bandwidth} is evaluated, as the key pool size needs to be positive at all times.}
    \label{fig:dynamic_pool_size_distance}
\end{figure}

After 10 hours of operation, a key reserve of more than 11 Gbits is formed at 50 km, falling to approximately 1 Gbit at 105 km. The dynamic size of the key pool as a function of time can be calculated for different configurations, using \autoref{eq:sk_balance} for $d_0=0$. Following an intense key consumption scenario, \autoref{fig:dynamic_pool_size_distance} demonstrates the time evolution of the key pool as a function of time for indicative communication distances. The plots also display key generation and key consumption terms. This calculation is faithful to the actual scheme, as excess keys are stored in the key pool in real time. As a result, the corresponding key availability condition is given by \autoref{eq:target_bandwidth}, dictating that uninterrupted secure communication requires the current pool size to remain positive at all times.

The plots demonstrate that shorter distances exhibit more frequent key contributions, sustaining a positive dynamic pool and forming a considerable key reserve. However, with increasing distances, a dead time appears after starting the operation, during which the key reserve is not sufficient to sustain secure communication. As an example, $l=82$ km requires more than 45 minutes of operation before the dynamic pool size becomes consistently positive. During this interval, real-time encryption cannot occur. To mitigate this effect, countermeasure strategies need to be applied, such as introducing QKD lead time.  

\subsection{QKD lead time}
\label{sec:qkd_lead}

To fulfill the key availability condition, QKD operation might need to be initialized before starting secure data exchange. This allows to form an initial key reserve for balancing the key consumption rate. As previous operation contributes additional complexity to the system, it is important to ensure that the QKD lead time is minimized. A systematic way of directly determining the minimum lead time required in each scenario is needed, given the transmission distance and use case parameters.

\begin{figure}[h]
    \centering
    \includegraphics[width=0.55\linewidth]{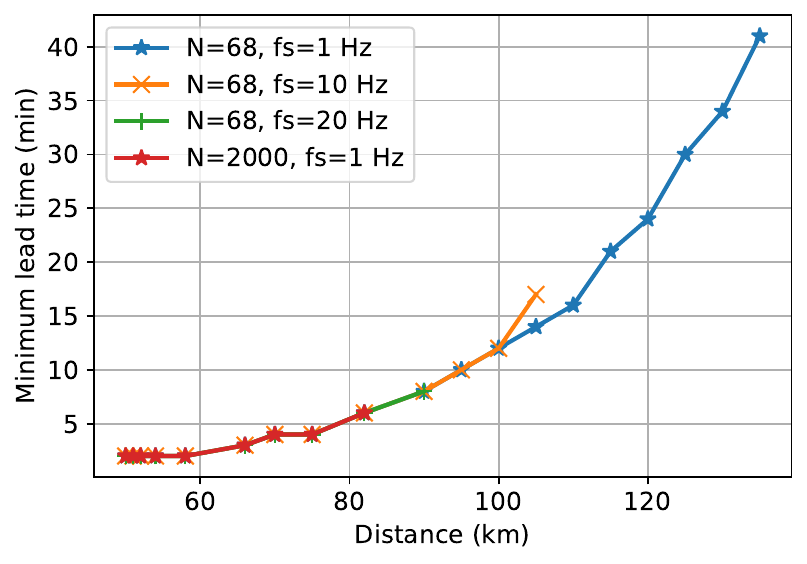}
    \caption{Minimum QKD lead times required for uninterrupted OTP encryption.}
    \label{fig:lead_times_v_distance}
\end{figure}

\begin{figure}[h]
    \centering
    \includegraphics[width=0.55\linewidth]{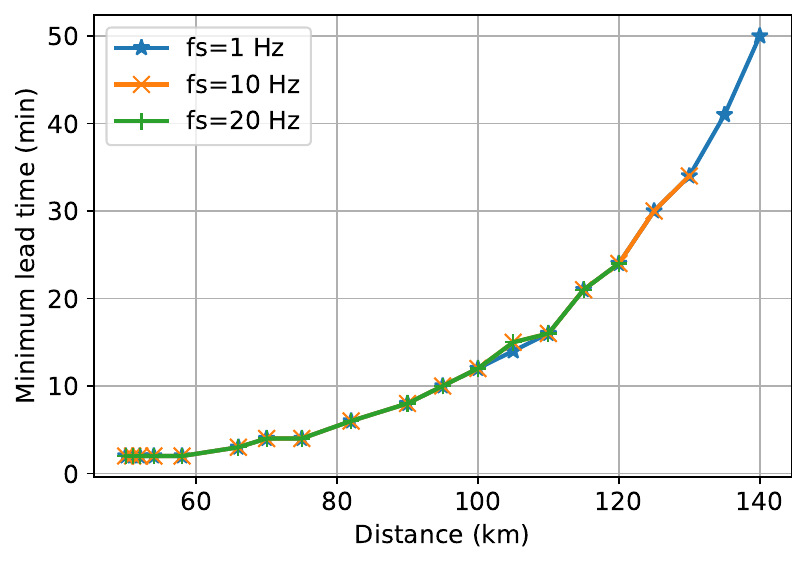}
    \caption{Minimum QKD lead time for AES-256 encryption with 128-bit IV under different parameter configurations. Key and IV are truly random bit sequences, updated in every encryption operation.}
    \label{fig:lead_times_v_distance_aes}
\end{figure}

For this reason, an automated script was developed which processes the QKD dataset and determines the lead time required for each use case configuration and distance. The algorithm evaluates \autoref{eq:sk_balance_w_lead_time} for $k \cdot \Delta \tau_{\text{ef}}=10$ hours of operation. The process is repeated for increasing lead time values until \autoref{eq:target_bandwidth} is satisfied.  In each case, the dynamic pool is evaluated using \autoref{eq:sk_balance_w_lead_time}. A configuration is considered nonviable if the obtained lead time exceeds 5 hours, suggesting that it should first be modified for bandwidth optimization. The script is executed for two signal sets ($N=68, N=2,000$), three sampling frequencies ($f_s=1$ Hz, $f_s=10$ Hz, $f_s=20$ Hz), two encryption algorithms (OTP, AES-256), and 21 distances, with results presented in \autoref{tab:headstarts_otp} and \autoref{tab:headstarts_aes256}. The trends corresponding to the remote monitoring use cases ($N=68$) and data historian use cases ($N=2,000$) are displayed in \autoref{fig:lead_times_v_distance} and \autoref{fig:lead_times_v_distance_aes}, for OTP and AES, respectively. 

In the remote monitoring scenario, a maximum distance of 135 km can be achieved when $f_s=1$ Hz. However, this comes at the cost of approximately 40 minutes of QKD lead time. If the frequency is increased to 20 cycles per second, the maximum distance is limited to 90 km, with a required lead time of 8 minutes. The results are indicative for the data historian use case. For $f_s=1$ Hz, the maximum achievable distance is 82 km for a lead time of 6 minutes. As the sampling rate increases, it is shown that OTP cannot be self-sustained. Although introducing prolonged lead times could be an option for supporting OTP with higher sampling rates for limited periods of time, it seems preferable to investigate data optimization methods instead. Such approach could lead to higher overall system reliability.

AES-256 is evaluated as a backup or alternative for use cases which combine intense data generation and long transmission distances. Lead time is not affected by the number of signals generated per time step as, unlike OPT, the block cipher defines a fixed key length not directly related to the size of plaintext. Due to the reduced key consumption rate, higher distances and sampling frequencies can be supported. The most demanding use case with 20 cycles/sec at 120 km requires only 24 minutes of QKD early operation, while distances up to 95 km require less than 10 minutes regardless of the data configuration. Notably, using AES practically expands the maximum achievable distance for data-heavy communication beyond 100 km. The AES-256 metrics confirm the potential to use the standard for those use cases where OTP cannot satisfy the key availability condition, by significantly reducing key consumption. This approach can be also found useful in case of an emergency or QKD system failure (e.g., fiber link interruption) for sustaining secure communication with increased key economy, explored in the next section.

\subsection{Key distribution failure}
\label{sec:post_failure}

The QKD experimental dataset is processed to determine the post-failure secure communication uptime $\Delta t_{\text{up}}$, for various parameter combinations. For each scenario evaluated, the optimal lead time $t_\text{{lead}}$ determined in \autoref{sec:qkd_lead} is assigned as the starting point of secure communication. In the first stage, QKD failure is assumed to occur at $t_\text{{failure}}=t_\text{{lead}}+1$ hour. The uptimes are determined by evaluating \autoref{eq:pf_condition_final2}.

\begin{figure}[h!]
    \centering
    % Subfigure 1
    \begin{subfigure}[b]{0.49\textwidth}
        \centering
        \includegraphics[width=\textwidth]{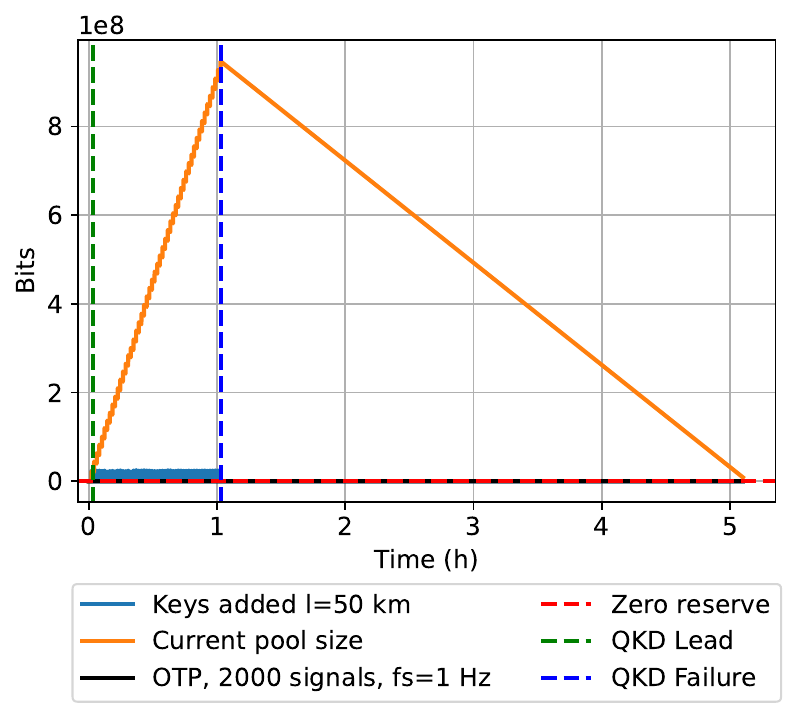} 
        \caption{$l=50$ km, $N=2000$, $f_s=1$ Hz}
        \label{fig:key_failure_50km}
    \end{subfigure}
    \hfill
    % Subfigure 2
    \begin{subfigure}[b]{0.49\textwidth}
        \centering
        \includegraphics[width=\textwidth]{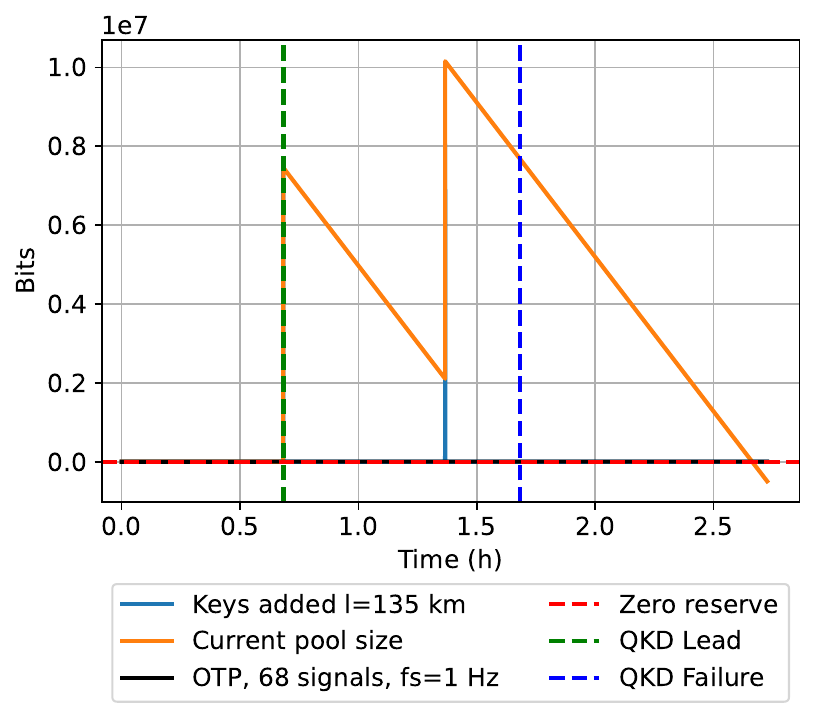} 
        \caption{$l=135$ km, $N=68$, $f_s=1$ Hz }
        \label{fig:key_failure_140km}
    \end{subfigure}
    
    \caption{Two representative use cases of key distribution failure. System lead time is assumed equal to the optimization value for each use case. Failure occurs 1 hour after secure communication is initiated. OTP is applied.}
    \label{fig:key_failure_50_140}
\end{figure}

The OTP results are shown in \autoref{tab:key_failure_otp}. The results demonstrate the dependence on distance and key consumption parameters. The maximum achievable distance for each configuration is still determined by the ability to obtain uninterrupted secure communication for a given lead time. For $l=50$ km the secure uptime is at least 4.1 hours for  $N=2,000$ (\autoref{fig:key_failure_50km}), reaching 149.1 hours when $N=68$. However, for 68 signals and $l=135$ km, secure operation is sustained for approximately 59 minutes before the key reserve is exhausted (\autoref{fig:key_failure_140km}).

\begin{figure}[h]
    \centering
    \includegraphics[width=0.65\linewidth]{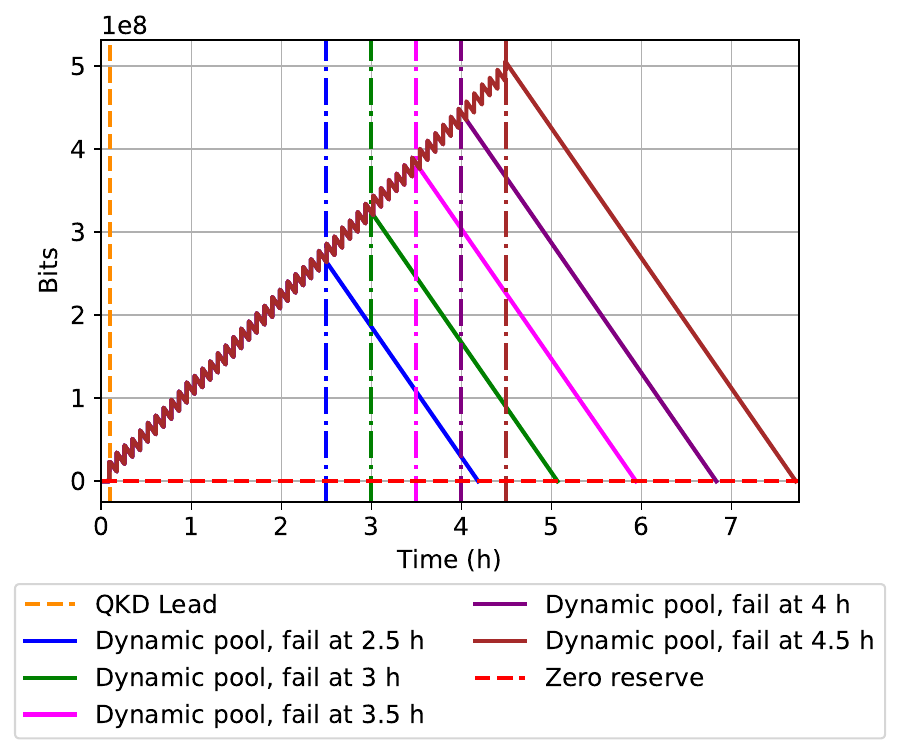}
    \caption{Dynamic key pool for QKD failure at different instances $t_{\text{fail}}$ ($l=82$ km, $N=68$, $f_s=20$ Hz, $p=32$ bits, $f_{\text{enc}}=1$, $t_{\text{lead}}=6$ min). }
    \label{fig:multi_failures}
\end{figure}

QKD failure time $t_{\text{fail}}$ can be arbitrarily set for investigating different events. In \autoref{fig:multi_failures}, the dynamic key pool at $l=82$ km is plotted for various $t_{\text{fail}}$ values. As a result, reversing the problem allows one to set a minimum acceptable time-to-failure as a target communication autonomy for an arbitrary I\&C system, and determine the required operational lead time.

The process is repeated with AES-256 instead of OTP (\autoref{tab:key_failure_aes}). Applying AES drastically increases the encryption uptime in case of QKD failure. At 1 Hz and distances less than 58 km, the system provides approximately 850 hours of encrypted data transmission. Although autonomy is reduced for higher rates and increasing distances, the system preserves computational security for at least 30 minutes, even in the worst-case scenario.

\begin{figure}[h]
    \centering
    \includegraphics[width=0.6\linewidth]{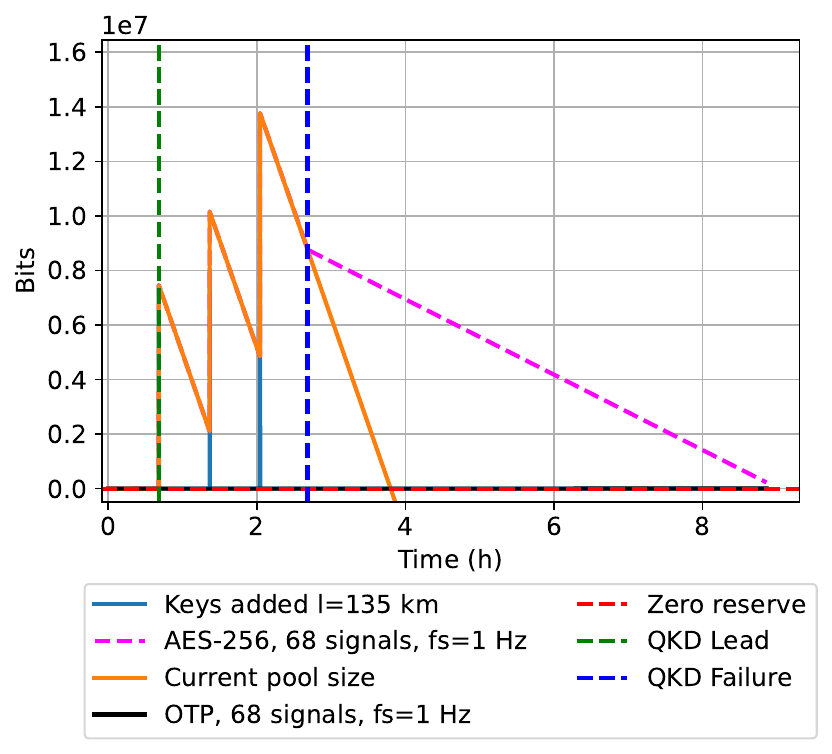}
    \caption{Dynamic key pool for at $l=135$ km. Switching to AES encryption after QKD failure provides an additional  5.2 hours of operation compared to OTP ($t_{\text{fail}}=t_{\text{lead}}+2$).}
    \label{fig:switch_to_aes}
\end{figure}

The identified trade-off between encryption robustness and key consumption allows to design a redundancy mechanism, where AES acts as a backup scheme until QKD communication is restored. This scenario can be particularly useful in larger distances with limited key generation, as it reduces key consumption and prolongs communication autonomy.  In \autoref{fig:switch_to_aes}, we recreate the key failure event at $l=135$ km with $N=68$ and $f_s=1$ Hz. QKD failure takes place 2 hours after secure communication begins. The plot displays the two response scenarios, maintaining OTP or switching to AES. While OTP provides an uptime of 67 minutes after failure, switching to AES after failure yields 6.3 hours. Similarly, for $t_{\text{fail}}=t_{\text{lead}}+1$ h, OTP offers 58 minutes of autonomy compared to 5.5 hours offered by switching to AES. The list of secure uptimes when switching from OTP to AES-256 at the time of failure is featured in \autoref{tab:otp_to_aes_switch}.

\section{Latency}
\label{sec:latency}

Latency is the second metric of interest for evaluating the performance of a use case. In this stage, the results from real-time secure communication cycles are studied. Secure communication between the two terminals (\autoref{fig:toshiba_schematic}) occurs following the procedure stages of \autoref{fig:data_com_loop}. The experiments are conducted for three cryptographic algorithm families (OTP, AES, ASCON). Cryptographic operations have been implemented through software, using open-source Python libraries. Time measurements are obtained for each use case configuration, for each of which the latency condition of \autoref{eq:latency_condition_final} is evaluated. The following sections present the measurement results.

\begin{figure}[h!]
    \centering
    % Subfigure 1
    \begin{subfigure}[b]{0.49\textwidth}
        \centering
        \includegraphics[width=\textwidth]{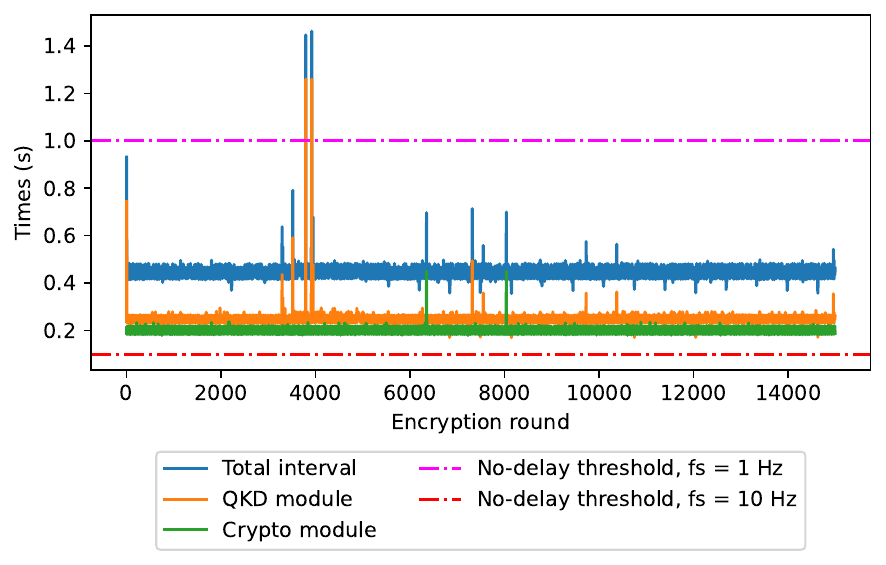} % Replace with your image
        \caption{OTP, $N=68$ signals, $f_{\text{enc}}=100\%$}
        \label{fig:latencies_otp:68}
    \end{subfigure}
    \hfill
    % Subfigure 2
    \begin{subfigure}[b]{0.49\textwidth}
        \centering
        \includegraphics[width=\textwidth]{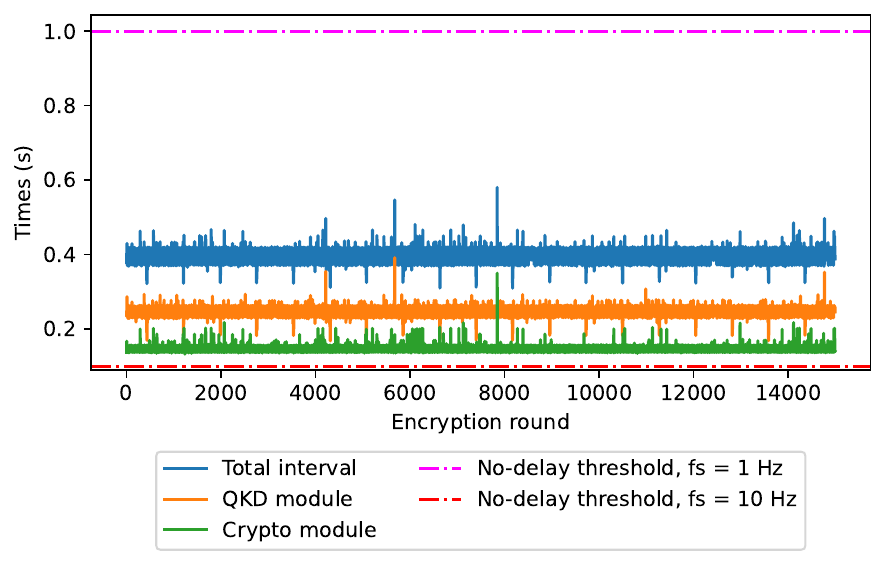} % Replace with your image
        \caption{OTP, $N=2000$ signals, $f_{\text{enc}}=100\%$}
        \label{fig:latencies_otp:2000}
    \end{subfigure}
        \begin{subfigure}[b]{0.49\textwidth}
        \centering
        \includegraphics[width=\textwidth]{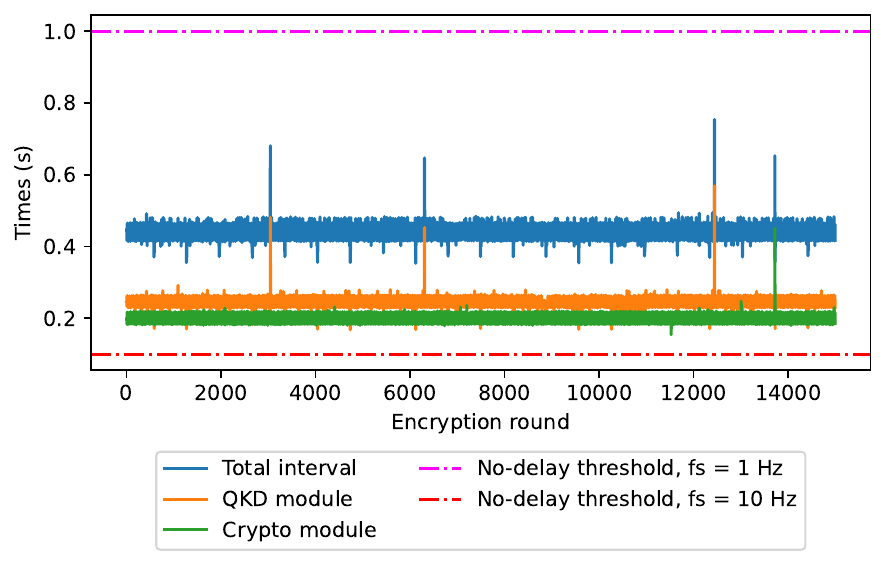} % Replace with your image
        \caption{AES-256, $N=68$ signals, $f_{\text{enc}}=17\%$}
        \label{fig:latencies_aes:68}
    \end{subfigure}
    \hfill
    % Subfigure 2
    \begin{subfigure}[b]{0.49\textwidth}
        \centering
        \includegraphics[width=\textwidth]{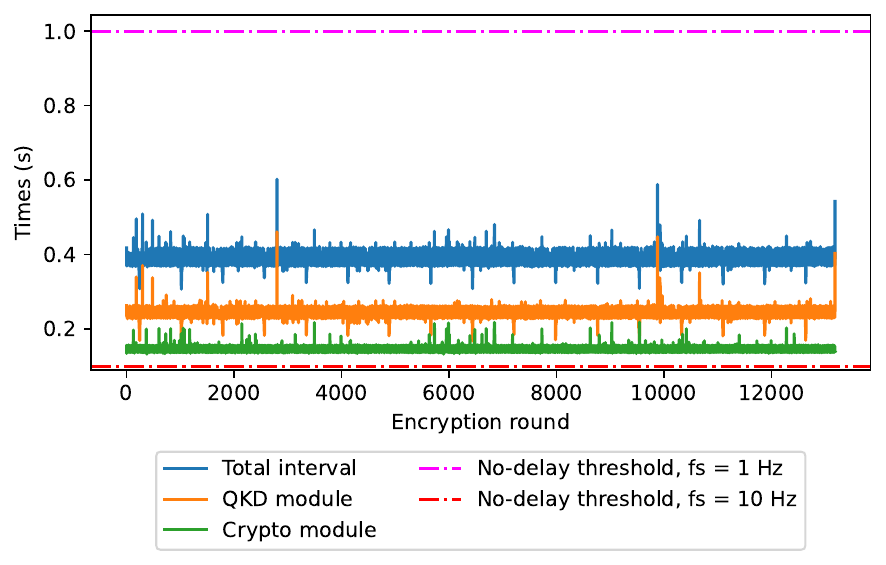} % Replace with your image
        \caption{AES-256, $N=2,000$ signals, $f_{\text{enc}}=0.6\%$}
        \label{fig:latencies_aes:2000}
    \end{subfigure}
    
    \caption{Latency for OTP and AES-256 encryption.  }
    \label{fig:latencies_otp}
\end{figure}

In the fundamental case, OTP encryption is considered. The reporting rate is identical to the sampling rate (i.e., each row of parameters corresponding to the current time step is encrypted and transmitted by itself). The 2,000 signals are fetched, encrypted and transmitted according to the communication procedure. The experiment is conducted for 15,000 data generation cycles, at a rate of 10 samples/sec. \autoref{fig:latencies_otp:2000} demonstrates the total latencies as well as those corresponding to the specific modules. The average time per data cycle is $ 395 \pm 15$ ms, out of which 248 ms and 145 ms are reported in the QKD and Crypto modules, respectively. The experiment is repeated only for the 68 core signals, with the results shown in \autoref{fig:latencies_otp:68}. Although there is significant difference in terms of key availability, the number of signals does not considerably affect latency in OTP encryption. Key distribution takes on average the same time as in the 2,000 signals case (248 ms), and the Crypto module requires approximately 50 ms more. The slight increase could be attributed to the overall higher efficiency of the encryption/decryption processes when handling larger data chunks.

AES-256 is the most secure variant of the Advanced Encryption Standard, featuring a 256-bit key and 128-bit IV. Since both the key and the IV are updated per data cycle, a total of 384 bits are requested at every iteration. Using \autoref{eq:key_reusability_abstract}, the key reusability factor is found equal to 17\% and 0.6\% for $N=68$ and $N=2,000$, respectively. Similarly to OTP, the latency metrics are evaluated and shown in \autoref{fig:latencies_aes:68} and \autoref{fig:latencies_aes:2000}. Although AES has higher complexity compared to the bitwise operations in OTP, the results do not indicate additional delay. Specifically, the average time of cryptographic operations is almost identical to the OTP cases (195 ms  for $N=68$ and 145 ms for $N=2,000$). The QKD module time is slightly reduced (5 ms), though the reduction is not proportional to the key reusability factor decrease. Therefore, the performance of AES in terms of latency is similar to OTP, with its primary advantage remaining higher key economy.

A similar approach is followed to evaluate the performance of ASCON variants. ASCON is the official selection of NIST for the novel field of Light-Weight Cryptography (LWC). LWC algorithms are designed around simpler operations, to enable cryptographic capabilities even in devices with limited computational resources \cite{turan_status_2023,gkouliaras_evaluating_2024}. In this implementation, three prominent variants are examined (ASCON-128, ASCON-128a, ASCON-80pq). Notably, ASCON-80pq is a post-quantum cryptography cipher, claiming to be more robust against a quantum computer-initiated attack \cite{nguyen_lightweight_2024}. 

\begin{figure}[h!]
    \centering
    % Subfigure 1
    \begin{subfigure}[b]{0.49\textwidth}
        \centering
        \includegraphics[width=\textwidth]{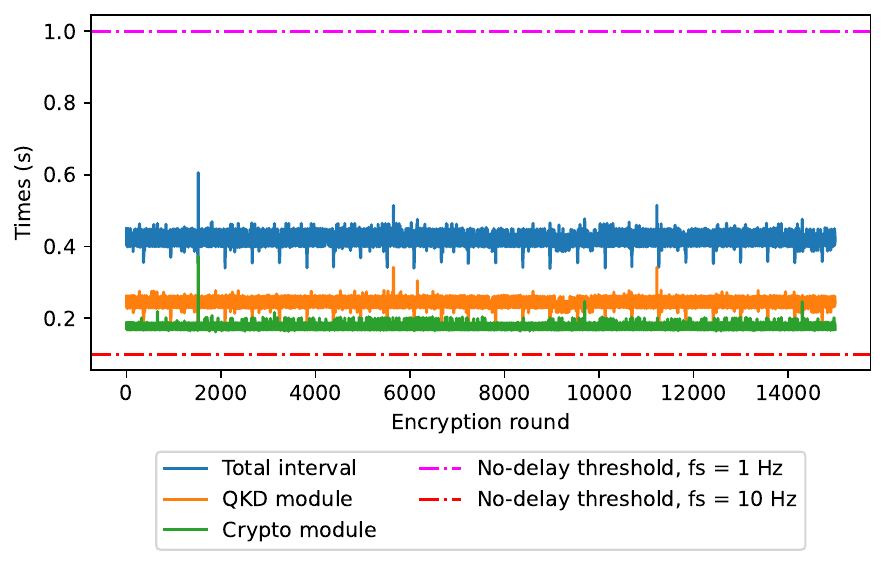} % Replace with your image
        \caption{ASCON-128, $f_{\text{enc}}=0.04\%$}
        \label{fig:latencies_ascon128}
    \end{subfigure}
    \hfill
    % Subfigure 2
    \begin{subfigure}[b]{0.49\textwidth}
        \centering
        \includegraphics[width=\textwidth]{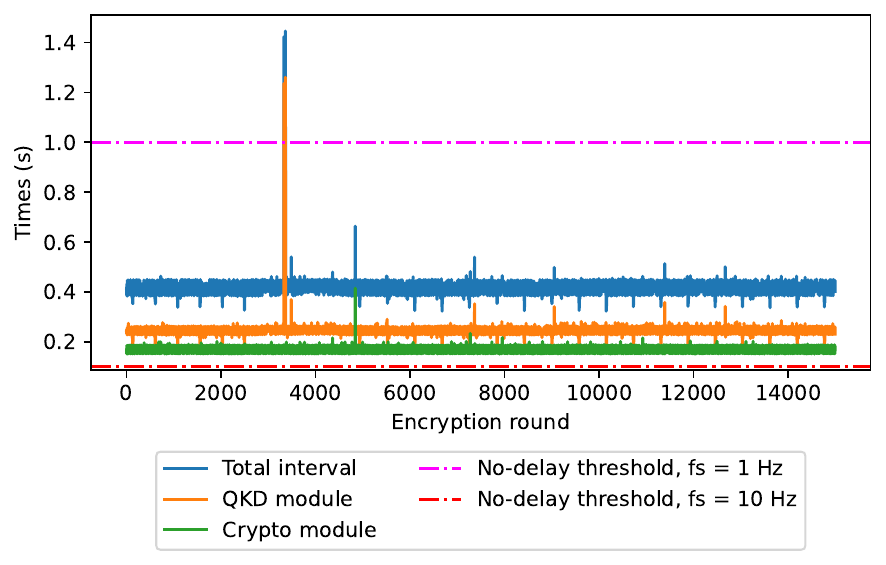} % Replace with your image
        \caption{ASCON-128a, $f_{\text{enc}}=0.04\%$}
        \label{fig:latencies_ascon128a}
    \end{subfigure}

    \begin{subfigure}[b]{0.49\textwidth}
        \centering
        \includegraphics[width=\textwidth]{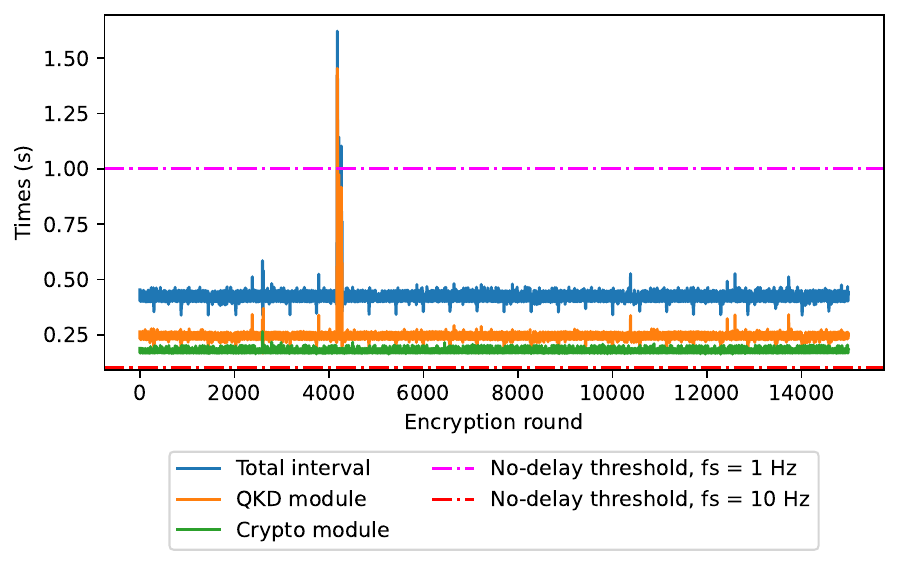} % Replace with your image
        \caption{ASCON-80pq,  $f_{\text{enc}}=0.045\%$}
        \label{fig:latencies_ascon80pq}
    \end{subfigure}
    
    \caption{Latencies in ASCON variants ($N=2000$ signals, $p=32$ bits)}
    \label{fig:latencies_ascon}
\end{figure}

In addition to the low computational resource requirements, the ASCON family offers authenticated encryption with associated data (AEAD). The key size is 128 bits for the 128 and 128a variants, while 80pq handles an 160-bit key. All three variants further require an 128-bit nonce. As in the previous cases, the key and nonce are refreshed per data generation cycle. Therefore, the key reusability factor is 0.4 \% for ASCON-128/128a and 0.45 \% for ASCON-80pq, when $N=2000$. The performance metrics are demonstrated in \autoref{fig:latencies_ascon}.

All three variants remain within the boundaries of 1 and 10 sampling iterations per second. While there is still delay for a use case of 10 Hz, the cryptographic module time is slightly reduced compared to OTP and AES by approximately 20-25 ms per cryptographic cycle. The QKD module intervals continue to average around 245 ms, confirming that the elapsed time for establishing communication between QKD servers and secure applications includes an unavoidable overhead.

\begin{figure}[h!]
    \centering
    \includegraphics[width=0.9\linewidth]
    {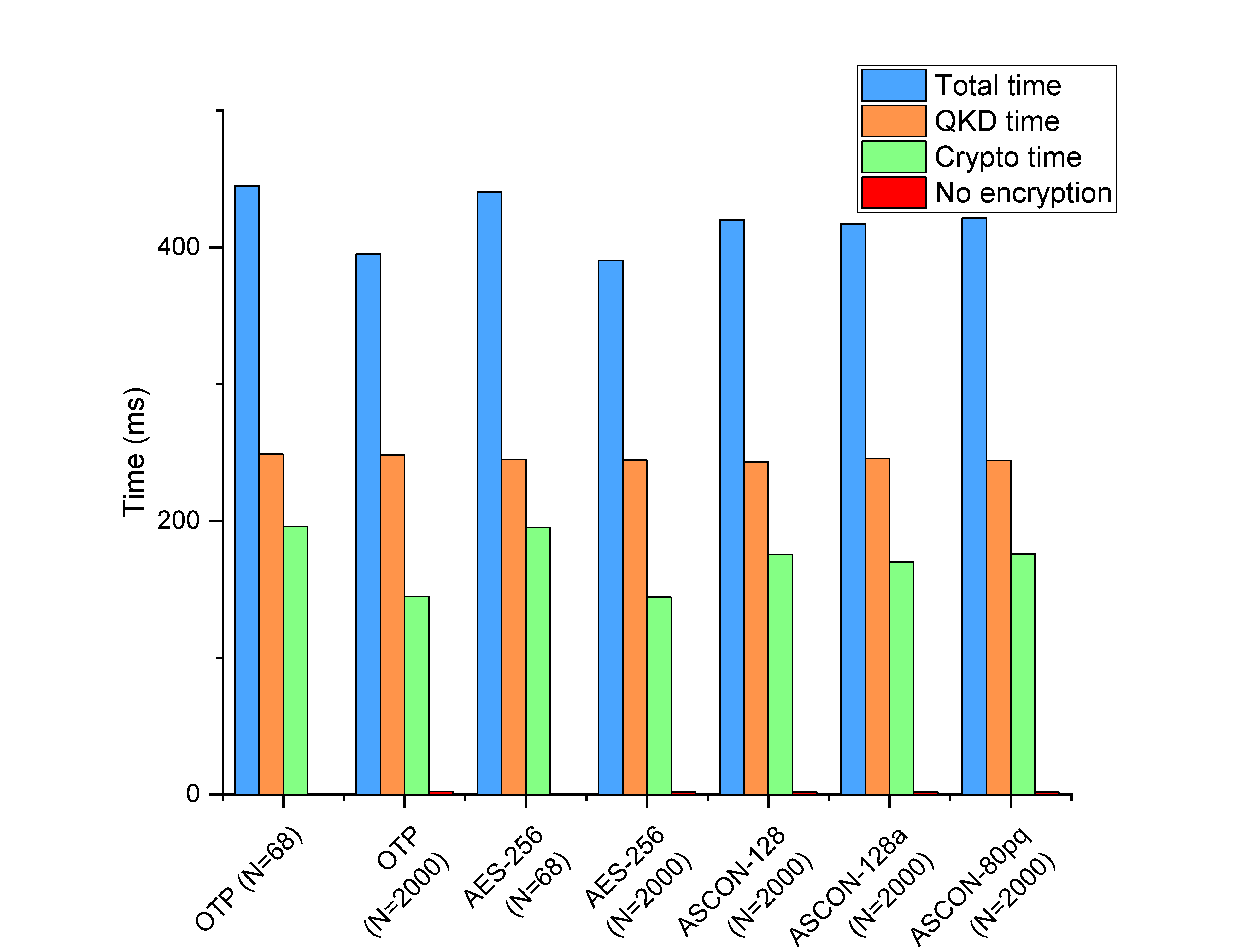}
    
    %{graphics/latency_bars.png}
    \caption{Average latency for evaluated cryptographic variants.}
    \label{fig:latency_bars}
\end{figure}

The obtained metrics for all tested algorithms and configurations are summarized in \autoref{tab:latency_fundamental_results} and displayed in \autoref{fig:latency_bars}. Overall, the total latency is consistently within the acceptable bounds, for the use case where data is sampled every second. In fact, there is an average margin of more than 0.5 seconds to immediately compensate for any delay outliers (e.g., the two peaks of \autoref{fig:latencies_otp:68}). The results confirm the potential of unconditionally secure remote operation, since $f_s=1$ Hz provides sufficient time resolution for the majority of monitored parameters.

\begin{table}[h!]
    \centering
    \caption{Latency metrics for different encryption algorithm variants and number of signals. Reporting frequency is equal to the sampling rate, therefore each operation corresponds to a single data row.}
        \centering
        \begin{tabular}{|c|c|c|c|c|c|c|c|}
            \hline 
            \multirow{2}{*}{Algorithm} & \multirow{2}{*}{Signals} & \multicolumn{2}{|c|}{Total time (ms)} & \multicolumn{2}{|c|}{QKD time (ms)} & \multicolumn{2}{|c|}{Crypto time (ms)} \\
            \cline{3-8}
            & & Mean & STD & Mean & STD & Mean & STD \\
            \hline 
            OTP & 68 & 444.95 & 20.91 & 248.76 & 17.28 & 195.79 & 11.47 \\
            \hline 
            OTP & 2,000 & 395.23 & 14.61 & 248.18 & 11.32 & 144.72 & 8.33 \\
            \hline 
            AES-256 & 68 & 440.47 & 16.66 & 244.79 & 11.70 & 195.32 & 10.83 \\
            \hline 
            AES-256 & 2,000 & 390.45 & 14.71 & 244.32 & 11.48 & 144.27 & 7.86 \\
            \hline 
            ASCON-128 & 2,000 & 419.91 & 14.49 & 243.05 & 11.16 & 175.28 & 8.00 \\
            \hline 
            ASCON-128a & 2,000 & 417.23 & 20.43 & 245.70 & 16.42 & 169.93 & 10.42 \\
            \hline 
            ASCON-80pq & 2,000 & 421.53 & 27.83 & 244.02 & 26.11 & 175.89 & 8.30 \\
            \hline
        \end{tabular}
        \label{tab:latency_fundamental_results}
    \end{table}

\section{Conclusion}
\label{sec:conclusion}

This paper presented an experimental demonstration of Quantum Key Distribution in a nuclear power reactor. An experimental setup was formed, leveraging PUR-1 research reactor and the commercial Toshiba QKD LD system. A secure communication model was formulated, defining detailed constraint conditions for evaluating the feasibility of a specific system configuration. 

The compatibility of QKD with nuclear reactor remote monitoring was confirmed through the design and replication of various use cases, characterized by different transmission lengths, reactor signals, sampling rates, and encryption algorithms. Secure, delay-free monitoring was achieved for up to 135 km for typical I\&C sampling rates.  The experimental data were further processed to generate key distribution failure scenarios, and subsequently determine the minimum QKD lead time required to prolong encrypted data exchange in case of an emergency. Results demonstrated that OTP-encrypted exchange can be achieved for 2,000 signals at up to 82 km when $f_s=1$ Hz. If a core of 68 signals is transmitted instead, the distance can reach 135 km, 105 km, and 90 km, for $f_s=1$ Hz, $f_s=10$ Hz, and $f_s=20$ Hz, respectively. Switching to AES-256, the maximum distance is expanded to 140 km for all 2,000 signals when $f_s=1$ Hz. 

Regarding future work, it is intended to investigate options for further minimizing latency, to enable secure transmission of higher data rates. Hardware implementation of the discussed encryption algorithms will be explored, in an attempt to further minimize latencies from the cryptographic module. In addition, procedure parallelization will be considered, to examine the potential of reducing the variance of latency attributed to QKD key requests.

\FloatBarrier

\section*{Acknowledgment}

This research is being performed using funding received from the DOE Office of Nuclear Energy’s Nuclear Energy University Program under contract DE-NE00009174. The authors would like to thank Dr.~Ben Cipiti at Sandia National Laboratories, and Katya LeBlanc at Idaho National Laboratories for fruitful discussions and expert input. Konstantinos Gkouliaras would like to acknowledge the Greek Atomic
Energy Commission for supporting his doctoral studies through a graduate fellowship.

\bibliographystyle{IEEEtran}
\bibliography{references_}

\newpage
\appendix
\section*{Appendix A: Result tables }

\begin{table}[h]
    \centering
    \caption{Minimum QKD operational lead time to secure key availability condition with OTP encryption for varying distance, number of signals and sampling frequency. Single precision ($p=32$ bits/value). Times determined from running the optimization algorithm on QKD data. }
    \begin{tabular}{|c|c|c|c|c|c|c|c|}
    \hline
    \multirow{3}{*}{\shortstack{Distance\\(km)}} & \multicolumn{6}{c|}{Minimum Time (min)} \\ \cline{2-7}
     & \multicolumn{3}{c|}{$N=68$} & \multicolumn{3}{c|}{$N=2000$} \\ \cline{2-7}
     & $f_s=1$ Hz & $f_s=10$ Hz & $f_s=20$ Hz & $f_s=1$ Hz & $f_s=10$ Hz & $f_s=20$ Hz \\ \hline
     \hline\hline
    50 & 2 & 2 & 2 & 2 & - & - \\ \hline
    51 & 2 & 2 & 2 & 2 & - & - \\ \hline
    52 & 2 & 2 & 2 & 2 & - & - \\ \hline
    54 & 2 & 2 & 2 & 2 & - & - \\ \hline
    58 & 2 & 2 & 2 & 2 & - & - \\ \hline
    66 & 3 & 3 & 3 & 3 & - & - \\ \hline
    70 & 4 & 4 & 4 & 4 & - & - \\ \hline
    75 & 4 & 4 & 4 & 4 & - & - \\ \hline
    82 & 6 & 6 & 6 & 6 & - & - \\ \hline
    90 & 8 & 8 & 8 & - & - & - \\ \hline
    95 & 10 & 10 & - & - & - & - \\ \hline
    100 & 12 & 12 & - & - & - & - \\ \hline
    105 & 14 & 17 & - & - & - & - \\ \hline
    110 & 16 & - & - & - & - & - \\ \hline
    115 & 21 & - & - & - & - & - \\ \hline
    120 & 24 & - & - & - & - & - \\ \hline
    125 & 30 & - & - & - & - & - \\ \hline
    130 & 34 & - & - & - & - & - \\ \hline
    135 & 41 & - & - & - & - & - \\ \hline
    140 & - & - & - & - & - & - \\ \hline
    145 & - & - & - & - & - & - \\ \hline
    \end{tabular}
    \label{tab:headstarts_otp}
    \end{table}

\begin{table}[h]
    \centering
    \caption{Minimum QKD operational lead time to secure key availability condition (AES-256 with  128-bit IV, $p=32$ bits)}
    \label{tab:headstarts_aes256}
    \begin{tabular}{|c|c|c|c|}
    \hline \multirow{2}{*}{\shortstack{Distance\\(km)}}  & \multicolumn{3}{|c|}{ Minimum lead time (minutes) } \\
    \cline{2-4} & ${f_s}=1 \mathrm{~Hz}$ & ${f_s}=10 \mathrm{~Hz}$ & ${f_s}=20 \mathrm{~Hz}$ \\
    \hline
    \hline
    \hline 50 & 2 & 2 & 2 \\
    \hline 51 & 2 & 2 & 2 \\
    \hline 52 & 2 & 2 & 2 \\
    \hline 54 & 2 & 2 & 2 \\
    \hline 58 & 2 & 2 & 2 \\
    \hline 66 & 3 & 3 & 3 \\
    \hline 70 & 4 & 4 & 4 \\
    \hline 75 & 4 & 4 & 4 \\
    \hline 82 & 6 & 6 & 6 \\
    \hline 90 & 8 & 8 & 8 \\
    \hline 95 & 10 & 10 & 10 \\
    \hline 100 & 12 & 12 & 12 \\
    \hline 105 & 14 & 15 & 15 \\
    \hline 110 & 16 & 16 & 16 \\
    \hline 115 & 21 & 21 & 21 \\
    \hline 120 & 24 & 24 & 24 \\
    \hline 125 & 30 & 30 & - \\
    \hline 130 & 34 & 34 & - \\
    \hline 135 & 41 & - & - \\
    \hline 140 & 50 & - & - \\
    \hline 145 & - & - & - \\
    \hline
    \end{tabular}
\end{table}

\begin{table}[hb]
    \centering
    \caption{System uptimes post QKD failure for various configurations. QKD lead time assigned in each configuration is the optimal value according to \autoref{tab:headstarts_otp}. QKD failure occurs with one-hour difference from the beginning of secure communication $t_{\text{fail}}=t_{\text{lead}}+1 h$. (OTP,  $p=32$ bits).}
    \begin{tabular}{|c|c|c|c|c|c|c|}
    \hline
    \multirow{3}{*}{\shortstack{Distance\\(km)}} & \multicolumn{6}{c|}{Time to failure (hours)} \\ \cline{2-7}
     & \multicolumn{3}{c|}{$N=68$} & \multicolumn{3}{c|}{$N=2000$} \\ \cline{2-7}
     & $f_s=1$ Hz & $f_s=10$ Hz & $f_s=20$ Hz & $f_s=1$ Hz & $f_s=10$ Hz & $f_s=20$ Hz \\ \hline
     \hline
    50  & 149.1311  & 14.0131  & 6.5066   & 4.1044  & -  & -  \\ 
    51  & 149.5491  & 14.0549  & 6.5275  & 4.1186   & -  & -  \\ 
    52  & 141.9219  & 13.2922  & 6.1461  & 3.8592  & -  & -  \\ 
    54  & 145.8739  & 13.6874  & 6.3437  & 3.9936  & -  & -  \\ 
    58  & 107.0347  & 9.8035  & 4.4017  & 2.6731  & -  & -  \\ 
    66  & 70.7764  & 6.1776  & 2.5888  & 1.4403  & -    & -  \\ 
    70  & 62.0944  & 5.3094  & 2.1547  & 1.1450  & -  & -  \\ 
    75  & 47.3700  & 3.8370  & 1.4185  & 0.6444  & -  & -  \\ 
    82  & 33.9222  & 2.4922  & 0.7461  & 0.1872  & -  & -  \\ 
    90  & 21.0992  & 1.2099  & 0.1050  & -  & -  & -\\ 
    95  & 15.8936  & 0.6893  & -  & -  & -  & -  \\ 
    100 & 13.0036  & 0.4004  & -  & -  & -  & -  \\ 
    105 & 10.9525     & 0.1953     & -  & -     & -  & -         \\ 
    110 & 7.1519  & -  & -  & -  & -  & -         \\ 
    115 & 7.0978  & -  & -  & -  & -         & -         \\ 
    120 & 4.2144  & -  & -  & -  & -         & -         \\ 
    125 & 3.0753  & -     & -  & -    & -         & -         \\ 
    130 & 1.5119  & -  & -  & -  & -         & -         \\ 
    135 & 0.9794  & -   & -  & -         & -         & -         \\ 
    140 & -  & -         & -         & -         & -         & -         \\ 
    145 & -         & -         & -         & -         & -         & -         \\ 
    \hline
    \end{tabular}
    \label{tab:key_failure_otp}
\end{table}

\begin{table}[b]
    \centering
    \caption{System uptimes post QKD failure for various configurations.  QKD lead time assigned in each configuration is the optimal value according to \autoref{tab:headstarts_aes256}. QKD failure occurs with one-hour difference from the beginning of secure communication (AES-256,  $p=32$ bits).}
    \begin{tabular}{|c|c|c|c|}
    \hline
    %\multirow{2}{*}{Distance (km)} & $f_s=1$ Hz & $f_s=10$ Hz & $f_s=20$ Hz \\ 
    \multirow{2}{*}{Distance (km)} & \multicolumn{3}{c|}{Time to failure (hours)} \\ \cline{2-4}
   &$f_s=1$ Hz & $f_s=10$ Hz & $f_s=20$ Hz \\
    \hline
    50  & $849.7436$    & 84.0744  & 41.5372  \\ 
    51  & $852.1128$    & 84.3113   & 41.6556  \\ 
    52  & $808.8911$    & 79.9891  & 39.4946  \\ 
    54  & $831.2864$    & 82.2286  & 40.6143  \\ 
    58  & $611.1981$    & 60.2198  & 29.6099  \\ 
    66  & 405.7328  & 39.6733  & 19.3366     \\ 
    70  & 356.5353  & 34.7535  & 16.8768  \\ 
    75  & 273.0969  & 26.4097  & 12.7048  \\ 
    82  & 196.8928  & 18.7893  & 8.8946  \\ 
    90  & 124.2297  & 11.5230  & 5.2615  \\ 
    95  & 94.7317  & 8.5731  & 3.7866  \\ 
    100 & 78.3547   & 6.9355  & 2.9677  \\ 
    105 & 66.7317  & 5.7732     & 2.3866  \\ 
    110 & 45.1953  & 3.6196  & 1.3098  \\ 
    115 & 44.8875  & 3.5887  & 1.2944         \\ 
    120 & 28.5483  & 1.9549  & 0.4774        \\ 
    125 & 22.0944  & 1.3094    & -         \\ 
    130 & 13.2350  & 0.4235  & -         \\ 
    135 & 10.2181     & -         & -         \\ 
    140 & 0.9214  & -         & -         \\ 
    145 & -         & -         & -         \\ 
    \hline
    \end{tabular}
    \label{tab:key_failure_aes}
\end{table}

\begin{table}[h]
\centering
\caption{System uptimes post QKD failure for various configurations.  QKD lead time assigned in each configuration is the optimal value according to \autoref{tab:headstarts_otp}. QKD failure occurs with one-hour difference from the beginning of secure communication. Encryption algorithm switches from OTP to AES-256 at the time of QKD failure.}

\begin{tabular}{|c|c|c|c|c|}
\hline
\multirow{3}{*}{\shortstack{Distance\\(km)}} &      \multicolumn{4}{c|}{Time to failure (hours)} \\ \cline{2-5}
   &$f_s=1$ Hz & $f_s=10$ Hz & $f_s=20$ Hz & $f_s$=1 Hz\\&  ($N=68$) &  ($N=68$) & ($N=68$) & ($N=2,000$) \\
\hline
50  & 845.0769  & 79.4077  & 36.8705  & 684.0769  \\
51  & 847.4461  & 79.6446  & 36.9890  & 686.4461  \\
52  & 804.2244  & 75.3224  & 34.8279  & 643.2244  \\
54  & 826.6197  & 77.5620  & 35.9477  & 665.6197  \\
58  & 606.5314  & 55.5531  & 24.9432  & 445.5314  \\
66  & 401.0661  & 35.0066  & 14.6700  & 240.0661  \\
70  & 351.8686  & 30.0869  & 12.2101  & 190.8686  \\
75  & 268.4303  & 21.7430  & 8.0382   & 107.4303  \\
82  & 192.2261  & 14.1226  & 4.2280   & 31.2261   \\
90  & 119.5631  & 6.8563   & 0.5948   & -         \\
95  & 90.0650   & 3.9063   & -        & -         \\
100 & 73.6881   & 2.2688   & -        & -         \\
105 & 62.0650   & 1.1067   & -        & -         \\
110 & 40.5286   & -        & -        & -         \\
115 & 40.2208   & -        & -        & -         \\
120 & 23.8817   & -        & -        & -         \\
125 & 17.4278   & -        & -        & -         \\
130 & 8.5683    & -        & -        & -         \\
135 & 5.5514    & -        & -        & -         \\
140 & -         & -        & -        & -         \\
145 & -         & -        & -        & -         \\
\hline
\end{tabular}
\label{tab:otp_to_aes_switch}
\end{table}

\end{document}